\documentclass[seceq]{ptptex}

\usepackage{graphicx}
\usepackage{url}

\notypesetlogo                       

\title{Explosive Nucleosynthesis in Magnetohydrodynamical Jets from Collapsars. II 
}

\subtitle{Heavy-Element Nucleosynthesis of {\it s}, {\it p}, {\it r}-Processes}   

\author{
Masaomi \textsc{Ono},$^{1,2,\,}$\footnote{E-mail: ono@yukawa.kyoto-u.ac.jp} 
Masa-aki \textsc{Hashimoto},$^2$ 
Shin-ichiro \textsc{Fujimoto},$^3$ 
Kei \textsc{Kotake}$^4$ 
and Shoichi \textsc{Yamada}$^5$
}

\inst{
$^1$Yukawa Institute for Theoretical Physics, Kyoto University, Kyoto 606-8502, Japan \\
$^2$Department of Physics, Kyushu University, Fukuoka 812-8581, Japan \\
$^3$Kumamoto National College of Technology, Kumamoto 861-1102, Japan \\
$^4$Division of Theoretical Astronomy/Center for Computational Astrophysics, National Astronomical Observatory of Japan, 
Mitaka 181-8588, Japan\\
$^5$Advanced Research Institute for Science and Engineering, Waseda University,
Tokyo 113-0033, Japan
}

\recdate{March 29, 2012; Revised August 4, 2012}

\abst{
We investigate the nucleosynthesis in a massive star of 70 $M_{\odot}$ with solar metallicity in the main sequence stage. 
The helium core mass after hydrogen burning corresponds to 32 $M_{\odot}$. Nucleosynthesis 
calculations have been performed during the stellar evolution and 
the jetlike 
supernova explosion of a collapsar model. We focus on the production of elements heavier than
 iron group nuclei. 
Nucleosynthesis calculations have been 
accomplished consistently from hydrostatic to dynamic stages 
by using large nuclear reaction networks, where 
the weak {\it s}-, {\it p}-, and {\it r}-processes are taken into account. 
We confirm that {\it s}-elements of $60<A<90$ are highly overproduced relative to the solar abundances in 
the hydrostatic nucleosynthesis. 
During oxygen burning, {\it p}-elements of $A>90$ are 
produced via photodisintegrations of seed {\it s}-elements. However, the produced {\it p}-elements are 
disintegrated in later stages except for $^{180}$Ta. 
In the explosive nucleosynthesis, elements of $90<A<160$ are significantly overproduced relative to the 
solar values owing to the {\it r}-process, which is very different from the results of spherical explosion models. 
Only heavy {\it p}-elements ($N>50$) are overproduced via the {\it p}-process 
because of the low peak temperatures in the 
oxygen- and neon-rich layers. 
Compared with the previous study of {\it r}-process nucleosynthesis calculations in the collapsar model of 40 $M_{\odot}$ by 
Fujimoto et al., 
[S.~Fujimoto, M.~Hashimoto, K.~Kotake and S.~Yamada, Astrophys.~J. \textbf{656} (2007), 382; 
S.~Fujimoto, N.~Nishimura and M.~Hashimoto, Astrophys.~J. \textbf{680} (2008), 1350], 
our jet model cannot contribute to the  third peak of the solar {\it r}-elements and 
intermediate {\it p}-elements, which have been much produced because of the distribution of the 
lowest part of electron fraction in the ejecta. 
Averaging the overproduction factors over the progenitor masses with 
the use of Salpeter's IMF, we suggest that the 
70 $M_{\odot}$ star could contribute to the solar weak {\it s}-elements of $60<A<90$ and neutron-rich elements of $90<A<160$. 
We confirm the 
primary synthesis of light {\it p}-elements in the ejected matter of high peak 
temperature. The ejected matter has [Sr/Eu] $\sim -0.4$, which is different from that of a typical 
{\it r}-process-enriched star CS22892-052 ([Sr/Eu] $\sim -1$). We find that Sr-Y-Zr isotopes are primarily synthesized in 
the explosive nucleosynthesis in a similar process of the primary production of light {\it p}-elements, which 
has been considered as 
one of the sites of a lighter element primary process (LEPP).
}

\begin{document}

\maketitle
\section{Introduction}

The origin 
of elements, particularly those heavier than iron, is still under debate.\cite{ref:thielemann_2011} 
Since charged particle reactions are difficult to produce those elements inside stars because of 
coulomb barriers, 
other nucleosynthesis processes, that is, two neutron capture processes, are required. 
One is the {\it r} (rapid)-process and the other is the {\it s} (slow)-process.\cite{ref:bbfh_1957} 
In the {\it r} ({\it s}) -process, neutron captures are faster (slower) than beta decays. \
Since the {\it r}-process requires high neutron exposure relative to seeds, 
the {\it r}-process favors low electron fraction ($Y_e$) and/or relatively 
high-entropy environments.\cite{ref:hoffman_1997} 

One of the promising sites of the {\it r}-process has been thought to be the neutrino-driven wind.
\cite{ref:hoffman_1997,ref:otsuki_2000,ref:wanajo_2007,ref:kuroda_2008} 
However, recent one-dimensional hydrodynamical simulations of the neutrino-driven wind 
with Boltzmann neutrino transport have revealed~\cite{ref:fischer_2010} that the electron fraction of the wind 
becomes high ($Y_e \gtrsim$ 0.5) and the entropy 
becomes low for the {\it r}-process. 
Therefore, other astrophysical sites such as neutron star mergers~\cite{ref:metzger_2010,ref:roberts_2011,ref:goriely_2011} or black hole winds~\cite{ref:wanajo_2011} have been proposed. 
However, the properties of ejecta such as densities, temperatures, 
and electron fractions are highly uncertain except for those in Ref.~\citen{ref:goriely_2011}, 
which makes even the qualitative analysis of the {\it r}-process difficult. 

In general, the {\it s}-process occurs at the end of core helium burning in massive stars and/or in AGB stars. 
Elements heavier than iron of $A<90$ are produced in massive stars, 
which is called the weak component of the {\it s}-process (weak {\it s}-process).\cite{ref:kappeler_2011} 
On the other hand, 
elements of $90<A<208$ are produced in AGB stars called the main component 
(main {\it s}-process).\cite{ref:kappeler_2011} 
The {\it s}-process is very sensitive to cross sections of neutron captures and $\beta$-decay rates, especially 
at the branching points such as $^{79}$Se and $^{85}$Kr.\cite{ref:kappeler_2011} 
Therefore, the nucleosynthesis in massive stars with the use of recent experimental cross sections 
is worth investigating. 
From the view point of astrophysics, part of the elements synthesized by the weak {\it s}-process 
could be ejected through the subsequent supernova explosion, where the produced {\it s}-elements should be 
the seeds of the {\it p}-process.\cite{ref:rayet_1995}

Abundances of metal-poor stars provide a good opportunity for understanding the nucleosynthesis because 
the abundances reflect the outcome only from a small number of supernova explosions. 
Observations of metal-poor stars have strongly suggested~\cite{ref:roederer_2010} 
that {\it r}-elements of the extremely metal-poor stars that have [Eu/Fe]\footnote{We adopt the usual notation 
[A/B] = log($N_{\rm A}$/$N_{\rm A}$) $-$ log($N_{\rm A}$/$N_{\rm A})_{\odot}$ for elements A and B.}  
$\gtrsim1$ (hereafter referred to as {\it r}-process-rich stars) have a ``universal'' abundance pattern, 
which reproduces the pattern of the solar system {\it r}-process 
abundances for $Z>56$. 
However, the abundances of $Z<56$ are not the case 
(e.g., Ref.~\citen{ref:sneden_2003}), and the observed [Sr, Y, Zr/Ba, Eu] 
ratios have dispersion in low-metallicity stars.\cite{ref:travaglio_2004} 
In particular, on the basis of the models of the chemical evolution of galaxies, 
it has been suggested~\cite{ref:travaglio_2004} that 
the abundances of Sr-Y-Zr ($Z=38,\, 39$, and 40, respectively) 
estimated from the contributions of the {\it s}- and {\it r}-processes are 
about 10 to 20\% less than the solar system abundances. 
As a consequence, a primary component from massive stars is needed to explain 8\% of the solar abundance of Sr 
and 18\% of those of Y and Zr, which should require a so-called lighter element primary process (LEPP).
\cite{ref:travaglio_2004} 

On the other hand, the mechanism of core-collapse supernova explosions is still a topic of debate. 
Pushed by recent observations 
revealing the aspherical natures of supernovae,
\cite{ref:wang_2002,ref:tanaka_2007} 
multidimensional studies of core-collapse supernova explosions 
have been elaborately performed as described in 
reviews.
\cite{ref:woosley_2005,ref:kotake_2006_rev,ref:kotake_2011_rev,ref:kotake_2012_rev1,ref:kotake_2012_rev2} 
Recent two/three-dimensional neutrino-radiation hydrodynamic simulations 
have shown successful supernova explosions, although 
in some of the models, the explosion energies are relatively small (10$^{49}$--10$^{50}$ erg).
\cite{ref:marek_2009,ref:suwa_2010,ref:takiwaki_2011,ref:kuroda_2012} 
QCD phase transition with a mixed phase of quarks and hadrons has also been 
reported as another 
possible supernova explosion mechanism even though the explosion is assumed to be spherical.\cite{ref:fischer_2011} 
Magnetohydrodynamical (MHD) simulations with some approximate neutrino transport schemes have  shown~\cite{ref:kotake_2004,ref:sawai_2005,ref:shibata_2006,ref:suwa_2007,ref:takiwaki_2009} 
jetlike explosions under some specific combinations of initial parameters for a strong magnetic 
field and differentially rapid rotation. While neutron stars are expected to be
left after supernova explosions, it has been suggested that a star of more than 25 $M_{\odot}$ may collapse 
to a black hole (BH)~\cite{ref:heger_2003_massive_star}; an accretion disk is formed around 
the BH if the star has enough angular momentum before the collapse. 
This system could produce a relativistic jet of gamma-ray bursts 
(GRBs)~\cite{ref:macfadyen_woosley_1999} due to MHD effects 
and/or neutrino heating around the rotational axis,\cite{ref:harikae_2010,ref:Zalamea_2011} whose system is 
called a collapsar model.\cite{ref:woosley_1993} 
MHD simulations in the context of the collapsar model have shown the formation of jets
~\cite{ref:koide_2002,ref:proga_2003,ref:mizuno_2004,ref:fujimoto_2006,ref:nagataki_2007,ref:harikae_2009
,ref:nagataki_2009,ref:nagataki_2011} due to winding-up effects of the magnetic field or the Blandford-Znajek process.\cite{ref:blandford_1977} 

Nucleosynthesis calculations 
of the {\it r}-process 
with the use  of a collapsar model 
of 40 $M_{\odot}$ 
have been performed extensively by Fujimoto et al.,\cite{ref:fujimoto_2007,ref:fujimoto_2008} 
where it is shown that the {\it r}-process 
would operate inside the jets. 
%
Explosive nucleosynthesis in GRB jets has also been investigated.\cite{ref:nagataki_2003,ref:nagataki_2006} 
%
Recent nucleosynthesis calculations in a three-dimensional MHD supernova model have suggested that such supernovae 
could be the sources of the {\it r}-process elements in the early Galaxy.\cite{ref:winteler_2012}
However, in those calculations, 
the produced nuclei are limited to primary synthesized ones inside the jets and comparisons with the solar system 
abundances have been focused on elements heavier than iron group nuclei. 
Nucleosynthesis calculations in spherical supernova explosions and detailed hydrostatic ones of the progenitors 
have proved that elements of $20<A<90$ are co-overproduced relative to the solar system abundances.
\cite{ref:rauscher_2002} However, elements 
of $A>90$ are not overproduced except for some {\it p}-elements. 
Recently, explosive nucleosynthesis calculations for a 15 $M_{\odot}$ presupernova model~\cite{ref:woosley_1995} 
with the solar metallicity based on two-dimensional hydrodynamical simulations have been 
performed,\cite{ref:fujimoto_2011} 
in which neutrino-driven explosions are triggered by adjusting the core neutrino luminosity parametrically. 
They have concluded that the overproductions 
relative to the solar abundances are similar to the results of spherical explosion models.\cite{ref:rauscher_2002} 
In our previous paper (Paper I),\cite{ref:ono_2009} 
we performed explosive nucleosynthesis calculations inside the jetlike explosions for the collapsar of 
a 70 $M_{\odot}$ star with the solar metallicity. These calculations include hydrostatic nucleosynthesis 
using a nuclear reaction network, which has 464 nuclei (up to $^{94}$Kr). 
In the present paper, we revisit the nucleosynthesis inside the jetlike explosion of the collapsar model and 
hydrostatic one taking into account all of the weak {\it s}-, {\it r}-, and {\it p}-processes. 
This makes it possible for us to estimate the consistent abundances of the ejecta. 
In particular, we study whether the collapsar model could be the source of elements heavier than iron. 

In \S2, we present the hydrostatic nucleosynthesis during the evolution 
from the helium burning stage to the onset of the core collapse. 
In \S3, we briefly summarize the MHD explosion model for the explosive nucleosynthesis and show the results. 
Section 4 is devoted to a summary of the overall results. 
We give some discussions with respect to uncertainties 
and remarks about the connection between metal-poor stars and the possibility for LEPP.

\section{Hydrostatic nucleosynthesis}

In Paper I, we investigated the nucleosynthesis in a massive star of 32 $M_{\odot}$ 
helium core corresponding to a main sequence star of 70 $M_{\odot}$.\cite{ref:hashimoto_1995} 
We have used the evolutional tracks with a nuclear reaction network, 
which includes 464 nuclei (up to $^{94}$Kr).\cite{ref:ono_2009} 
In massive stars, the (weak) {\it s}-process should occur at the end of helium and carbon burning stages. 
However, in the previous paper, we have included only nuclei of $A<94$; we could not discuss 
the weak {\it s}-process. 
Therefore, in the present paper, we perform a more detailed nucleosynthesis calculation with 
a larger nuclear reaction network. 

\subsection{Stellar model, initial compositions, and physical inputs}


 A star of $M_{\rm ms}\sim70~M_\odot$ with the solar metallicity could correspond to the upper limit of 
accreting BH models (collapsars), because more massive stars suffer from the strong mass 
loss.\cite{ref:heger_2003_massive_star} 
As a result, the size of the helium core will be affected considerably by the mass loss.
However, the rate of mass loss is still very uncertain~\cite{ref:puls_2008}; we calculate the
evolution of a massive helium core, $M_\alpha = 32~M_\odot$, without the mass loss as an extreme case, 
which is worth studying to see the final fate for the series of helium core evolution. 

 
 We calculate the nucleosynthesis along each evolutional track of the Lagrange mass 
from the stage of gravitational contraction of the core to the initiation of iron core collapse. 
The calculation has been carried out by using changes in the density ($\rho$), temperature ($T$) and 
convective regions. This is the so-called postprocess nucleosynthesis calculation. 
In convective regions, elements are mixed and compositions become almost uniform. 
Therefore, the region 
is calculated 
as 
one zone with averaged mass fractions and nuclear reaction 
rates as in Ref.~\citen{ref:prantzos_1987}. 

Toward {\it s}-process calculation, we construct a new reaction network including 1714 nuclei up 
to $^{241}$U, in which the reaction rates are based on a new REACLIB compilation, 
namely, JINA REACLIB database~\cite{ref:cyburt_2010}. 
The included elements are given in Table \ref{table:nuclide_1714}. 
The experimental ($n$,\,$\gamma$) reaction rates in JINA REACLIB are based on KADoNiS projects.
\cite{ref:dillmann_2006} 
In stellar environments, $\beta$-decay rates could be different from the values in laboratories. 
Takahashi and Yokoi (hereafter referred to as TY87)~\cite{ref:takahashi_1987} calculated theoretically 
the $\beta$-decay and electron capture 
rates for elements heavier than $^{59}$Ni and tabulated the rates taking into account thermally enhanced 
ionized and excited states in stellar interiors. 
The table ranges over 
5$\times$10$^7$ $\leq T \leq $ 5$\times$10$^8$ K and 
10$^{26}$ $\leq n_e \leq $ 3$\times$10$^{27}$ cm$^{-3}$, 
where $n_e$ is the number density of electrons. 
We adopt the temperature and density dependence of the rates if available. 
Unfortunately, the range of the table is limited only for $T$ and $\rho$
of the helium burning stage. 
If the temperature and density are outside of the table, we use the values of the edges of the table. 
Reaction rates concerning $^{180}$Ta are specially treated as noted in Appendix A, 
because $^{180}$Ta has a long-lived isomeric state. 
As suggested by Prantzos et al.\cite{ref:prantzos_1987}, the timescales of neutron-induced reactions 
($\approx$ $10^{-4}$ s) are much shorter than those of the variation of abundances of 
the other elements ($\approx 10^9$--$10^{11}$ s) as well as convective timescales, 
and thereby, neutron abundance is determined locally under thermal equilibrium conditions. 
This indicates that neutrons are not mixed uniformly in convective regions. 
Therefore, to calculate abundances in a convective region as one zone including the effects of different 
neutron abundances over the region, 
we adopt doubly averaged reaction rates for ``neutron-induced'' reactions of 
 ($n$,\,$\gamma$), ($n$,\,$p$), and ($n$,\,$\alpha$), according to the same method described 
in Ref.~\citen{ref:prantzos_1987}. 



Let $X(i)$ denote the mass fraction of the element $i$. 
The initial mass fractions are assumed to be $X(^{4}$He$)=0.981$ and 
$X(^{14}$\rm N$)=0.0137$, where all the original CNO isotopes are assumed 
to be converted to $^{14}$N during the core hydrogen burning. 
Mass fractions of the heavier elements are taken to be proportional to the solar system abundances~\cite{ref:anders_1989} 
(e.g., $X(^{56}\rm Fe)=1.17\times10^{-3}$).

%
The evolutionary changes in composition with respect to the stellar structure 
such as $^{12}$C, $^{16}$O, and $^{20}$Ne 
are slightly different from the results 
with the use of the reaction network of 464 nuclei~\cite{ref:ono_2009} originating from the differences 
in the initial abundances and adopted reaction rates. 
The produced mass fractions of elements of $A>94$, 
which have not been included in the 464 network, amount to about 10$^{-6}$. 
Therefore, differences of 10$^{-6}$ in mass fractions may be introduced. 
Although there are some differences in the main composition as described in Paper I between the original 
stellar evolution model~\cite{ref:hashimoto_1995} 
and postprocess hydrostatic nucleosynthesis calculations with the reaction network of 464 or 1714 nuclei, the differences 
are not so large for our purpose. 
\begin{table}
\begin{center}
\caption{1714 nuclides contained in the nuclear reaction network for the hydrostatic nucleosynthesis.}
\label{table:nuclide_1714}
\renewcommand{\arraystretch}{1.1}
\begingroup
\small
\begin{tabular}{cccccccc}
\vspace{-0.2cm} \\
\hline\hline
Nuclide & $A$ & Nuclide & $A$ & Nuclide & $A$ & Nuclide & $A$\\
\hline
\multicolumn{1}{l}{H} & \multicolumn{1}{r|}{1 -- 3} & 
\multicolumn{1}{l}{Cr} & \multicolumn{1}{r|}{46 -- 62} & 
\multicolumn{1}{l}{Ag} & \multicolumn{1}{r|}{100 -- 123} &
\multicolumn{1}{l}{Yb} & \multicolumn{1}{r}{162 -- 184} \\
\multicolumn{1}{l}{He} & \multicolumn{1}{r|}{3 -- 6} & 
\multicolumn{1}{l}{Mn} & \multicolumn{1}{r|}{48 -- 65} & 
\multicolumn{1}{l}{Cd} & \multicolumn{1}{r|}{103 -- 126} & 
\multicolumn{1}{l}{Lu} & \multicolumn{1}{r}{165 -- 187} \\
\multicolumn{1}{l}{Li} & \multicolumn{1}{r|}{6 -- 9} & 
\multicolumn{1}{l}{Fe} & \multicolumn{1}{r|}{50 -- 68} & 
\multicolumn{1}{l}{In} & \multicolumn{1}{r|}{105 -- 129} &
\multicolumn{1}{l}{Hf} & \multicolumn{1}{r}{168 -- 189} \\
\multicolumn{1}{l}{Be} & \multicolumn{1}{r|}{7 -- 12} & 
\multicolumn{1}{l}{Co} & \multicolumn{1}{r|}{52 -- 70} &
\multicolumn{1}{l}{Sn} & \multicolumn{1}{r|}{108 -- 132} & 
\multicolumn{1}{l}{Ta} & \multicolumn{1}{r}{171 -- 191} \\
\multicolumn{1}{l}{B} & \multicolumn{1}{r|}{8 -- 14} & 
\multicolumn{1}{l}{Ni} & \multicolumn{1}{r|}{54 -- 73} & 
\multicolumn{1}{l}{Sb} & \multicolumn{1}{r|}{111 -- 135} &
\multicolumn{1}{l}{W} & \multicolumn{1}{r}{174 -- 194} \\
\multicolumn{1}{l}{C} & \multicolumn{1}{r|}{9 -- 18} & 
\multicolumn{1}{l}{Cu} & \multicolumn{1}{r|}{57 -- 75} & 
\multicolumn{1}{l}{Te} & \multicolumn{1}{r|}{113 -- 137} & 
\multicolumn{1}{l}{Re} & \multicolumn{1}{r}{177 -- 197} \\
\multicolumn{1}{l}{N} & \multicolumn{1}{r|}{11 -- 21} & 
\multicolumn{1}{l}{Zn} & \multicolumn{1}{r|}{59 -- 78} & 
\multicolumn{1}{l}{I} & \multicolumn{1}{r|}{117 -- 140} & 
\multicolumn{1}{l}{Os} & \multicolumn{1}{r}{180 -- 200} \\
\multicolumn{1}{l}{O} & \multicolumn{1}{r|}{13 -- 23} & 
\multicolumn{1}{l}{Ga} & \multicolumn{1}{r|}{62 -- 80} & 
\multicolumn{1}{l}{Xe} & \multicolumn{1}{r|}{120 -- 142} & 
\multicolumn{1}{l}{Ir} & \multicolumn{1}{r}{183 -- 203} \\
\multicolumn{1}{l}{F} & \multicolumn{1}{r|}{14 -- 26} & 
\multicolumn{1}{l}{Ge} & \multicolumn{1}{r|}{64 -- 83} & 
\multicolumn{1}{l}{Cs} & \multicolumn{1}{r|}{122 -- 145} & 
\multicolumn{1}{l}{Pt} & \multicolumn{1}{r}{186 -- 206} \\
\multicolumn{1}{l}{Ne} & \multicolumn{1}{r|}{17 -- 28} & 
\multicolumn{1}{l}{As} & \multicolumn{1}{r|}{67 -- 86} & 
\multicolumn{1}{l}{Ba} & \multicolumn{1}{r|}{125 -- 148} & 
\multicolumn{1}{l}{Au} & \multicolumn{1}{r}{188 -- 209} \\
\multicolumn{1}{l}{Na} & \multicolumn{1}{r|}{19 -- 30} & 
\multicolumn{1}{l}{Se} & \multicolumn{1}{r|}{69 -- 89} & 
\multicolumn{1}{l}{La} & \multicolumn{1}{r|}{128 -- 150} & 
\multicolumn{1}{l}{Hg} & \multicolumn{1}{r}{191 -- 212} \\
\multicolumn{1}{l}{Mg} & \multicolumn{1}{r|}{21 -- 33} & 
\multicolumn{1}{l}{Br} & \multicolumn{1}{r|}{72 -- 91} & 
\multicolumn{1}{l}{Ce} & \multicolumn{1}{r|}{131 -- 153} & 
\multicolumn{1}{l}{Tl} & \multicolumn{1}{r}{194 -- 215} \\
\multicolumn{1}{l}{Al} & \multicolumn{1}{r|}{23 -- 35} & 
\multicolumn{1}{l}{Kr} & \multicolumn{1}{r|}{74 -- 93} & 
\multicolumn{1}{l}{Pr} & \multicolumn{1}{r|}{133 -- 156} & 
\multicolumn{1}{l}{Pb} & \multicolumn{1}{r}{198 -- 217} \\
\multicolumn{1}{l}{Si} & \multicolumn{1}{r|}{25 -- 38} & 
\multicolumn{1}{l}{Rb} & \multicolumn{1}{r|}{76 -- 96} & 
\multicolumn{1}{l}{Nd} & \multicolumn{1}{r|}{136 -- 158} &
\multicolumn{1}{l}{Bi} & \multicolumn{1}{r}{202 -- 220} \\
\multicolumn{1}{l}{P} & \multicolumn{1}{r|}{27 -- 40} & 
\multicolumn{1}{l}{Sr} & \multicolumn{1}{r|}{79 -- 98} &
\multicolumn{1}{l}{Pm} & \multicolumn{1}{r|}{138 -- 160} & 
\multicolumn{1}{l}{Po} & \multicolumn{1}{r}{205 -- 222} \\
\multicolumn{1}{l}{S} & \multicolumn{1}{r|}{29 -- 42} & 
\multicolumn{1}{l}{Y} & \multicolumn{1}{r|}{81 -- 101} & 
\multicolumn{1}{l}{Sm} & \multicolumn{1}{r|}{141 -- 163} &
\multicolumn{1}{l}{At} & \multicolumn{1}{r}{209 -- 224} \\
\multicolumn{1}{l}{Cl} & \multicolumn{1}{r|}{31 -- 45} & 
\multicolumn{1}{l}{Zr} & \multicolumn{1}{r|}{83 -- 103} & 
\multicolumn{1}{l}{Eu} & \multicolumn{1}{r|}{143 -- 165} & 
\multicolumn{1}{l}{Rn} & \multicolumn{1}{r}{212 -- 227} \\
\multicolumn{1}{l}{Ar} & \multicolumn{1}{r|}{33 -- 48} & 
\multicolumn{1}{l}{Nb} & \multicolumn{1}{r|}{86 -- 106} & 
\multicolumn{1}{l}{Gd} & \multicolumn{1}{r|}{146 -- 168} & 
\multicolumn{1}{l}{Fr} & \multicolumn{1}{r}{215 -- 229} \\
\multicolumn{1}{l}{K} & \multicolumn{1}{r|}{35 -- 50} & 
\multicolumn{1}{l}{Mo} & \multicolumn{1}{r|}{89 -- 109} & 
\multicolumn{1}{l}{Tb} & \multicolumn{1}{r|}{148 -- 171} & 
\multicolumn{1}{l}{Ra} & \multicolumn{1}{r}{217 -- 232} \\
\multicolumn{1}{l}{Ca} & \multicolumn{1}{r|}{37 -- 53} & 
\multicolumn{1}{l}{Tc} & \multicolumn{1}{r|}{91 -- 112} & 
\multicolumn{1}{l}{Dy} & \multicolumn{1}{r|}{151 -- 174} & 
\multicolumn{1}{l}{Ac} & \multicolumn{1}{r}{222 -- 234} \\
\multicolumn{1}{l}{Sc} & \multicolumn{1}{r|}{39 -- 55} & 
\multicolumn{1}{l}{Ru} & \multicolumn{1}{r|}{93 -- 115} & 
\multicolumn{1}{l}{Ho} & \multicolumn{1}{r|}{154 -- 176} & 
\multicolumn{1}{l}{Th} & \multicolumn{1}{r}{225 -- 237} \\
\multicolumn{1}{l}{Ti} & \multicolumn{1}{r|}{41 -- 57} & 
\multicolumn{1}{l}{Rh} & \multicolumn{1}{r|}{96 -- 117} & 
\multicolumn{1}{l}{Er} & \multicolumn{1}{r|}{157 -- 179} & 
\multicolumn{1}{l}{Pa} & \multicolumn{1}{r}{227 -- 239} \\
\multicolumn{1}{l}{V} & \multicolumn{1}{r|}{43 -- 59} & 
\multicolumn{1}{l}{Pd} & \multicolumn{1}{r|}{98 -- 120} & 
\multicolumn{1}{l}{Tm} & \multicolumn{1}{r|}{160 -- 181} & 
\multicolumn{1}{l}{U} & \multicolumn{1}{r}{231 -- 241} \\
\hline
\end{tabular}
\endgroup
\end{center}
\end{table}

\subsection{{\it S}- and {\it p}-processes} 

\begin{figure}
\begin{center}
\includegraphics[width=13cm,keepaspectratio,clip]{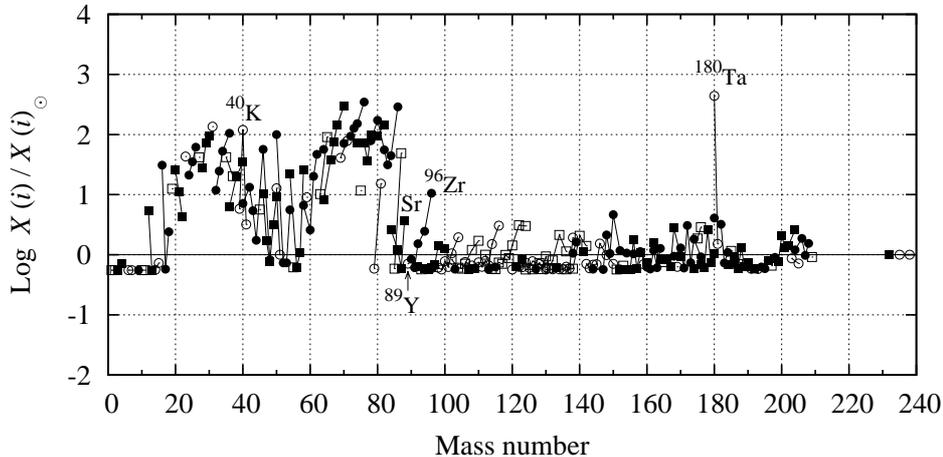}
\end{center}
\caption{Overproduction factors $X(i)/X(i)_{\,\odot}$ against the mass number $A$ 
averaged over the star including the hydrogen envelope at the beginning of the core collapse. 
Distinguished symbols connected by lines indicate the isotopes. 
}
\label{fig:abund_over_631}
\end{figure}

Let us focus on the production of elements heavier than iron and the 
weak {\it s}- and {\it p}-processes in the hydrostatic nucleosynthesis. 
Figure~\ref{fig:abund_over_631} shows overproduction factors $X(i)/X(i)_{\,\odot}$ at the beginning of the core collapse, 
where $X(i)$ is the value averaged over the star including the hydrogen-rich envelope of 38 $M_{\odot}$, and 
$X(i)_{\odot}$ is that of the solar system abundances.  
It is noted that in the estimation of the overproduction factors, 
we adopt the value of 4.55$\times$10$^{9}$ yr ago for $^{40}$K, which is 
a long-lived radioactive nucleus (the half-life is 1.25$\times$10$^9$ yr) and the present abundance is about one order of magnitude less 
than that when the solar system was born ($\sim$ 4.5 yr ago). There is inconsistency between the initial abundance of $^{40}$K 
and the value in the estimation of the overproduction factor. However, the overproduction of $^{40}$K is not determined by the initial 
abundance of $^{40}$K but by the abundance of $^{39}$K as described below. 
Overall, elements of $60<A<90$ are highly overproduced relative to the solar ones; the overproduction level ranges 
over 10$^2$ -- 10$^3$, which is similar to the weak {\it s}-process scenario proposed 
by many previous studies of the {\it s}-process in massive stars.
\cite{ref:prantzos_1987,ref:langer_1989,ref:rayet_2000} 

\begin{figure}
\begin{center}
\includegraphics[width=13cm,keepaspectratio,clip]{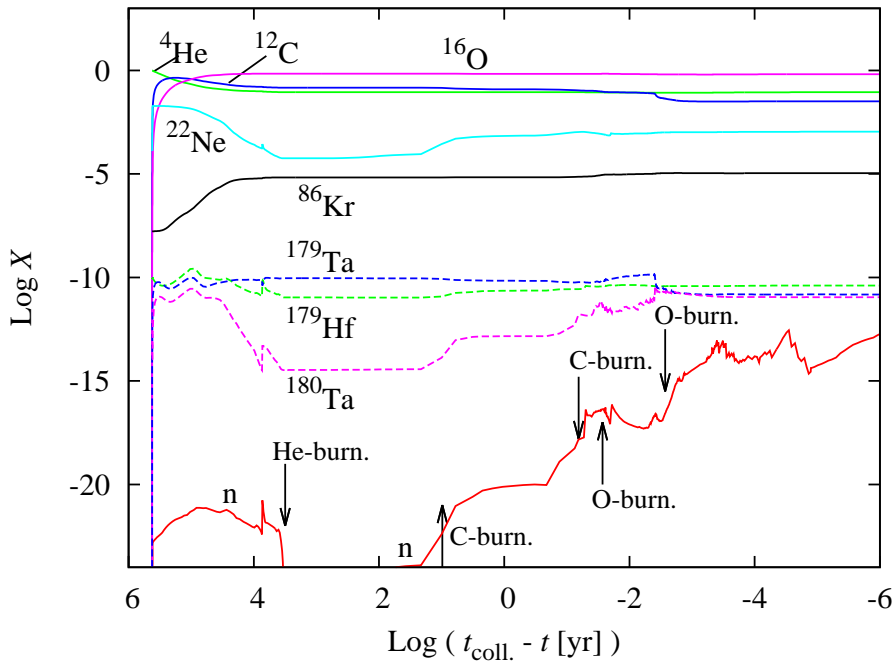}
\end{center}
\caption{Changes in mass fractions of selected elements averaged over the helium core against 
remaining time before the core collapse ($t_{\rm coll.}$ = 4.22$\times$10$^5$ yr). 
Upward (downward) arrows denote the time of the start (end) of the burning stages. 
}
\label{fig:frac_evol}
\end{figure}

Figure~\ref{fig:frac_evol} shows the time evolution of mass fractions of selected elements averaged over the whole star. 
$^{86}$Kr, which is one of the representative {\it s}-elements, is overproduced at the helium burning stage. 
We confirm that the overproduced elements of $60<A<90$ are mainly produced during the helium core burning, and the 
neutrons are mainly supplied by the $^{22}$Ne($\alpha$,\,$n$)$^{25}$Mg reaction (Fig.~\ref{fig:frac_evol}) 
as pointed out in previous studies (see Ref.~\citen{ref:kappeler_2011} for a recent review). 
$^{22}$Ne is produced by the sequence of 
$^{14}$N($\alpha$,\,$\gamma$)$^{18}$F($\beta^{-} \nu$)$^{18}$O($\alpha$,\,$\gamma$)$^{22}$Ne reactions. 
$^{40}$K is produced by the $^{39}$K($n$,\,$\gamma$)$^{40}$K reaction in the helium burning stage. 
The solar system abundances of $^{39}$K and $^{40}$K are 3516 and 0.44 (the present value) 
(normalized as the abundance of silicon to be 10$^6$),\cite{ref:anders_1989} respectively. 
Therefore, even if the small amount of $^{39}$K is converted to $^{40}$K,
we regard it to be much overproduced relative to the solar value. 
$^{180}$Ta is also overproduced in the helium burning stage by the sequence of 
$^{179}$Hf($\beta^{-}$)$^{179}$Ta($n$,\,$\gamma$)$^{180}$Ta reactions (see Fig.~\ref{fig:frac_evol}). 
It is noted that the reaction channel of $^{179}$Hf($\beta^{-}$)$^{179}$Ta is closed if we do not use 
the $\beta$-decay rates of TY87 
because $^{179}$Hf is stable in the laboratory. However, overproduced $^{180}$Ta decays from a thermally 
populated ground state at the end of the helium burning, and the overproduction level 
returns almost to the initial value (Fig.~\ref{fig:frac_evol}). 
%
In the helium burning stage, $^{96}$Zr is overproduced more than 10 times (Fig.~\ref{fig:abund_over_631}) 
relative to the solar value by neutron captures, which are faster than $\beta^{-}$-decays of 
$^{95}$Zr ($\tau_{1/2}\sim$ 64 d). However, Sr and Y are not so overproduced. 
%
Although the overproduction levels of the elements of $10<A<90$ do not increase significantly after the helium burning, 
some elements of $A>90$ are produced after that. 

In the carbon and neon burning stages, neutrons are mainly supplied by reactions of $^{12}$C($^{12}$C,\,$n$)$^{23}$Mg, 
$^{13}$C($\alpha$,\,$n$)$^{16}$O, $^{17}$O($\gamma$,\,$n$)$^{16}$O, and 
$^{25}$Mg($\alpha$,\,$n$)$^{28}$Si. 
The $^{22}$Ne($\alpha$,\,$n$)$^{25}$Mg reaction is also activated in the helium burning shell. 
Since $^{179}$Hf is produced by neutron captures, $^{180}$Ta is overproduced again relative to the solar value 
via the sequence of $^{179}$Hf($\beta^{-}$)$^{179}$Ta($n$,\,$\gamma$)$^{180}$Ta. 

\begin{figure}
\begin{center}
\includegraphics[width=12cm,keepaspectratio,clip]{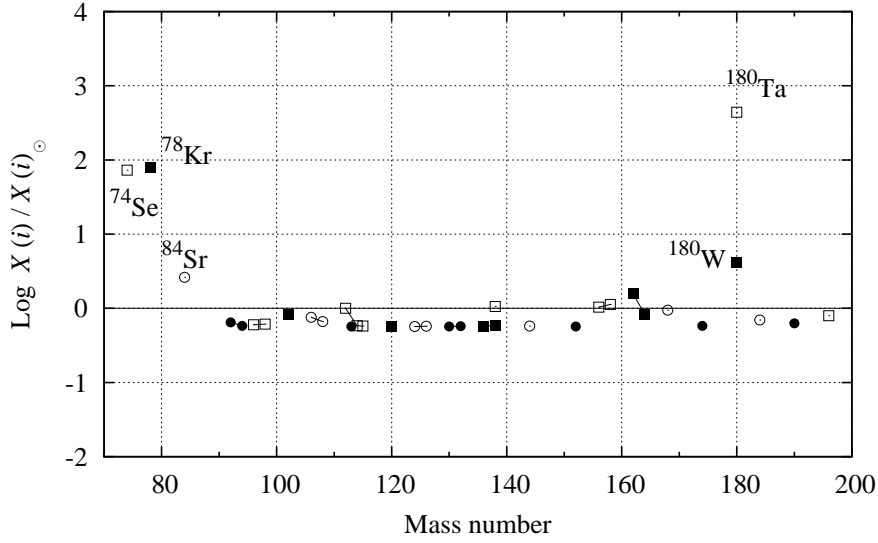}
\end{center}
\caption{Same as Fig.~\ref{fig:abund_over_631} but for 35 {\it p}-elements.}
\label{fig:p_nucl_631}
\end{figure}

During the oxygen burning stage, the temperature becomes high 
($T_{\rm c}\sim$ 2$\times$10$^{9}$ K, where $T_{\rm c}$ is the temperature at the center) and 
photodisintegrations of {\it s}-elements activate the {\it p}-process in the oxygen- and neon-rich layers. 
The possibility of the {\it p}-process in hydrostatic evolution of massive stars was proposed in previous studies.
\cite{ref:arnould_1976,ref:nomoto_1988,ref:hashimoto_1989,ref:rauscher_2002} 
We find that seed {\it s}-elements that have larger mass numbers tend to be disintegrated into {\it p}-elements 
by ($\gamma$,\,$n$) reactions, and light {\it p}-elements of $^{74}$Se, $^{78}$Kr, and $^{84}$Sr are produced 
at the oxygen- and neon-rich layers by ($\gamma$,\,$n$) and ($\gamma$,\,$p$) reactions (Fig.~\ref{fig:p_nucl_631}). 
It is noted that $^{180}$Ta is produced more and more by the $^{181}$Ta($\gamma$,\,$n$)$^{180}$Ta reaction. 
Since the solar abundances of $^{180}$Ta are much smaller than those of $^{181}$Ta, conversion of small amounts of 
$^{181}$Ta leads to the overproduction of $^{180}$Ta relative to the solar value. 

In the later stages, produced {\it p}-elements of $A>90$ are disintegrated by subsequent ($\gamma$,\,$n$) reactions and 
$\beta^{+}$-decays except for $^{180}$Ta. After all, {\it p}-elements whose overproduction factors 
are greater than 10 are only $^{74}$Se, $^{78}$Kr, and $^{180}$Ta at 
the beginning of the collapse (see Fig.~\ref{fig:p_nucl_631}). 
It is noted that ($n$,\,$\gamma$) and ($\gamma$,\,$n$) reactions are in thermal equilibrium at this stage. 

\begin{figure}
\begin{center}
\includegraphics[width=13cm,keepaspectratio,clip]{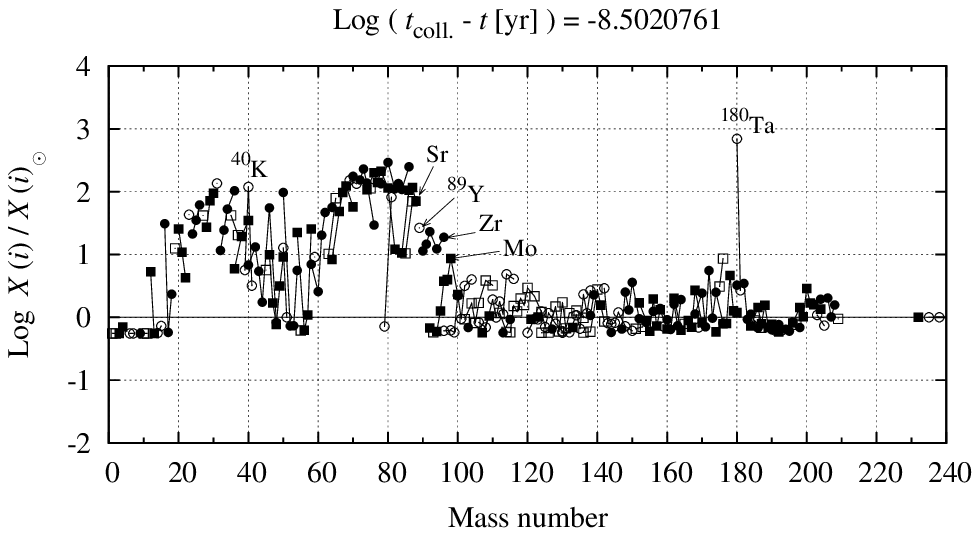}
\end{center}
\caption{
Same as Fig.~\ref{fig:abund_over_631} but for the results with use of laboratory $\beta$-decay rates
instead of those of TY87.
}
\label{fig:abund_over_TY87_off}
\end{figure}

The quantitative assessment of the adopted beta decay rates (TY87) should require more experimental 
data.\cite{ref:kappeler_2011}  
Moreover, the table of the $\beta$-decay rates of TY87 covers only 
the temperature and density of the helium burning stage as mentioned in \S2.1. Therefore, we have also calculated 
the nucleosynthesis using laboratory $\beta$-decay rates 
instead of those of TY87.  
Although overall 
overproduction levels change only by some factors as seen in Fig.~\ref{fig:abund_over_TY87_off}, 
those of more neutron-rich isotopes of 
$60<A<85$ 
tend to be decreased. 
We find that the production of elements for 
$85<A<100$ 
is very sensitive to the differences in 
$\beta$-decay rates, and Sr, Y, Zr, and Mo tend to be overproduced. 
Although the channel of $^{179}$Hf($\beta^{-}$)$^{179}$Ta($n$,\,$\gamma$)$^{180}$Ta is closed without TY87, 
$^{180}$Ta is produced by the $^{181}$Ta($\gamma$,\,$n$)$^{180}$Ta reaction after the carbon burning and 
remains through ($n$,\,$\gamma$) $\rightleftharpoons$ ($\gamma$,\,$n$) equilibrium in the later stages. 

Let us summarize the hydrostatic nucleosynthesis. 1) We confirm the weak {\it s}-process scenario: 
{\it s}-elements of $60<A<90$ are highly overproduced relative to the solar abundances. 
2) High overproductions of $^{180}$Ta could be attributed to the higher 
density and temperature evolutional tracks of the 70 $M_{\odot}$ star. 
3) The {\it p}-process in the oxygen- and neon-rich layers occurs after the carbon burning stage. 
However, the produced {\it p}-elements do not remain due to subsequent ($\gamma$,\,$n$) reactions and $\beta^{+}$-decays. 
We suggest that for smaller massive stars, the overproductions of $^{180}$Ta are significantly 
decreased because of the lower density and temperature evolutional paths. 
On the other hand, the {\it p}-process could become important 
because the produced {\it p}-elements of larger mass numbers may survive in 
low-density and low-temperature environments. 









\section{Explosive nucleosynthesis in a magnetohydrodynamical jet}

In Paper I, we investigated the explosive nucleosynthesis with the nuclear reaction 
network including 464 nuclei (up to $^{94}$Kr).\cite{ref:ono_2009} 
In the present paper, we recalculate the explosive nucleosynthesis with a much larger reaction network 
and focus on the production of heavier elements, 
that is, {\it p}- and {\it r}-elements. In this section, we present the review of the explosion model, the method 
of explosive nucleosynthesis, and the results combined with that of the hydrostatic nucleosynthesis. 

\subsection{Supernova explosion model}

In Paper I, we have constructed supernova explosion models 
using a collapsar model.\cite{ref:ono_2009} 
Here, we briefly summarize the explosion model for the nucleosynthesis calculation. 
We have performed two-dimensional MHD simulations of the collapsar model using a 
nonrelativistic MHD code, ZEUS-2D,\cite{ref:stone_1992a,ref:stone_1992b} 
which is modified~\cite{ref:kotake_2003} for handling supernova simulations 
with a realistic nuclear equation of state (EOS) based on the relativistic mean field theory.\cite{ref:shen_1998} 

For a low-density region of $\rho<10^5$ g cm$^{-3}$, another EOS is connected~\cite{ref:yasutake_2007} 
smoothly at the density boundary, which consists of the nonrelativistic ions, partially 
degenerate relativistic electrons, and radiation.

We have taken into account neutrino cooling by electron-positron ($e^{\pm }$) pair captures on nucleons, 
$e^{\pm }$ pair annihilation, and nucleon-nucleon bremsstrahlung. 
We neglect to include both
detailed neutrino transport and heating processes, because the maximum density 
remains less than 10$^{10}$ g cm$^{-3}$ in our calculations. We discuss the effects of neutrino absorptions on the nucleosynthesis in \S 3.2. 
BH was mimicked as an inner free absorption 
boundary and gravitational point source with pseudo Newtonian potential.\cite{ref:paczynsky_wiita_1980} 
We adopted the spherical coordinate ($r$,\,$\theta$,\,$\phi$), and the computation domain was taken 
from the inner boundaries $r_{\rm in} =$ 50 -- 200 km to 3$\times10^{4}$ km, which covers inner oxygen-rich layers. 

The initial presupernova model is the 32 $M_{\odot}$ helium core corresponding to an $M_{\rm ms}$ = 70 $M_{\odot}$ 
star,\cite{ref:hashimoto_1995} which is the same stellar evolution model obtained in the previous section. 
The initial configuration of angular velocity and magnetic field was implemented by analytical form 
with parameters as in the previous study.\cite{ref:mizuno_2004,ref:kotake_2004,ref:fujimoto_2006,ref:fujimoto_2008} 
The initial angular velocity is written as follows:
\begin{equation}
\Omega\,(r) = \Omega_{0} \frac{r^{2}_{0}}{r^{2} + r^{2}_{0}},
\end{equation}
where $r$ is the radius from the center, and $\Omega_0$ and $r_0$ are model parameters. 
The initial toroidal magnetic field is given in proportional to the angular velocity distribution as 
\begin{equation}
B_{\phi}\,(r) = B_{0} \frac{r^{2}_{0}}{r^{2} + r^{2}_{0}},
\end{equation}
where $B_0$ is a model parameter. 
We adopt the R51 model as in 
Paper I 
for the explosive nucleosynthesis, 
which has the largest amount of the mass end energy ejection rates among the investigated models. 
The parameters of the initial angular velocity and magnetic field are 
$\Omega_{0}$ = 5 s$^{-1}$, $r_{0}$ = 1500 km, $B_0 = 5.7 \times 10^{12}$ G, and $B_Z = 5 \times10^{11}$ G, 
where $B_Z$ is the initial uniform poloidal magnetic field along the rotational axis. 
The specified parameters of the rotation and magnetic field correspond to the model of the rapid rotation 
and the strong magnetic field. Recent stellar evolution models indicate~\cite{ref:heger_2005} that if the magnetic 
field is taken into account, the resultant specific angular momentum of the central region 
becomes smaller than that required for the typical collapsar model ($j \sim$ 10$^{17}$ cm$^2$ s$^{-1}$
~\cite{ref:macfadyen_woosley_1999}). 
In our simulation, the jet is triggered by the central magnetic pressure, which grows due to the compression and winding-up 
effects of the magnetic field, and the amplified magnetic field reaches around $\sim$ 10$^{15}$ G. If the magnetorotational instability (MRI)~\cite{ref:balbus_Hawley_1998} 
successfully operates in the core-collapse phase, the magnetic field could be quickly amplified to the same level 
from an initial magnetic field weaker than that ascribed in the present paper. 
However, resolving MRI in a global simulation is very hard and not feasible in the present calculations. 
Therefore, we assume that some mechanisms 
such as MRI amplify the magnetic field rapidly from a weak initial magnetic field and we mimic the situation 
by simply imposing a strong initial magnetic field. Note that the reached magnetic field strength $\sim$ 10$^{15}$ G 
is comparable to that at saturation due to MRI.\cite{ref:obergaulinger_2009} 

The resulting total ejection mass and explosion energy at the end of the simulation ($t_{f} = 1.504$ s) 
are 0.124 $M_{\odot}$ and 3.02$\times10^{50}$ erg, respectively. 
The specific angular momentum $j$ after the formation of a disklike structure is about 5 $\times$ 10$^{16}$ cm$^2$ s$^{-1}$ 
at the radius of 500 km. 
The disk extends to about 1000 km from the center at the accretion phase. 
After the strong jet formation, an expanding bow shock is generated at the outer region of the disk and 
the matter residing in the region begins 
to expand outward along the equatorial axis (see the description of Fig.~\ref{fig:particle} in \S 3.2). 
However, even after the formation of the jet and the expanding bow shock, 
the accretion continues at the inner edge of the disk and the accretion rate maintains the value of 
0.1 -- 1 $M_{\odot}$ s$^{-1}$ with a few factors declined in 1 s. 
Therefore, our model can be regarded as a collapsar model. 
It is noted that the jet 
obtained by the simulation 
is 
mildly relativistic ($\lesssim$ 0.1 $c$) and 
baryon rich 
($\gtrsim$ 0.1 $M_{\odot}$). 
In contrast, ultrarelativistic GRB jets should be baryon poor ($\sim$ 10$^{-5}$ $M_{\odot}$~\cite{ref:piran_2005}). 
Moreover, the event rate of mildly relativistic jets could be larger than those of normal GRB jets.\cite{ref:granot_2004} 
Therefore, both 
the ejected mass and event rate 
suggest that the contribution to the chemical evolution of galaxies 
could be large compared with that of ultrarelativistic ones. 




%



\subsection{Computational method of nucleosynthesis inside the jet}

\begin{figure}
\begin{minipage}{6.9cm}
\begin{center}
\includegraphics[width=6cm,keepaspectratio,clip]{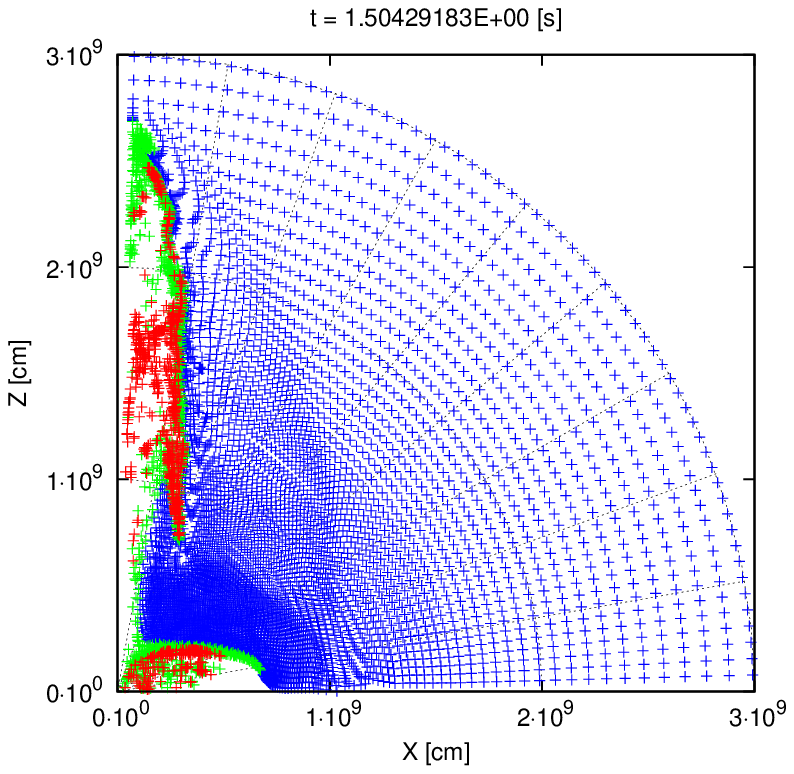}
\caption{Distribution of the tracer particles on the $XZ$-plane at the end of the simulation ($t_f$ = 1.504 s). 
The particles initially located at the inner iron core, 
Si-rich layers and oxygen-rich layers are indicated in red, green, and blue, respectively.
}
\label{fig:particle}
\end{center}
\end{minipage}
\hspace{0.2cm}
\begin{minipage}{6.9cm}

\includegraphics[width=7cm,keepaspectratio,clip]{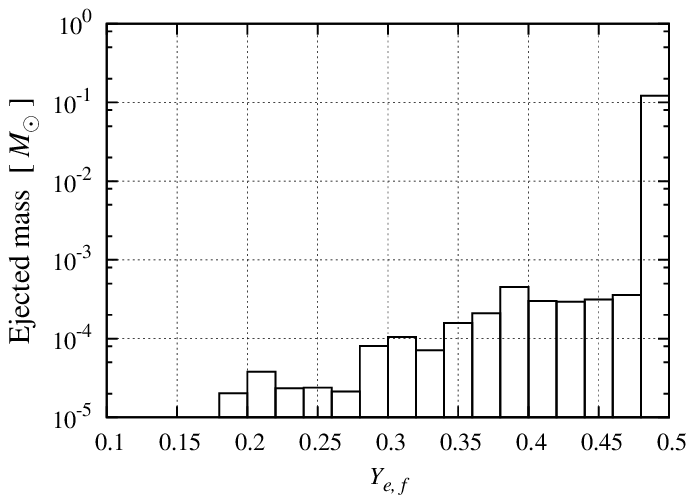}
\caption{Ejected masses against electron fraction at the end of NSE stage.
}
\label{fig:hist_ye}
\end{minipage}
\end{figure}

For calculating the nucleosynthesis inside the MHD jet, 20,000 tracer particles are distributed 
over the computational domain between 
1000 
km and $ 3 \times 10^{4}$~km from the center, 
which covers initially the region from around the iron core surface to the inner oxygen-rich layer. All the matter initially located 
at radii smaller than 
1000 
km is absorbed into the inner boundary. 
The Lagrange evolution of density and temperature of each tracer particle can be obtained from the method described in 
Refs.~\citen{ref:nagataki_1997} and \citen{ref:fujimoto_2007}, by which we calculate the nucleosynthesis and follow 
the change in composition. 
Figure~\ref{fig:particle} shows the distribution of the tracer particles at the end of the simulation 
($t_f$ = 1.504 s). 
The particles initially located at the inner iron core, 
Si-rich layers, and oxygen-rich layers are indicated in red, green, and blue, respectively. 
Some fractions of the particles initially located at the inner iron core are ejected by the jet. 
It should be noted that in Fig.~\ref{fig:particle}, we can see a ``blank" region in which there are no particles 
in the equatorial regions. 
The blank region corresponds to the expanding region after the jet formation as mentioned in \S3.1. 
The density of the equatorial blank region is $\rho \lesssim$ 10$^5$ g cm$^{-3}$. 
In the tracer particle method, 
the distributions of the particles tend to be sparse in expanding and low-density regions. We can also see relatively sparse regions 
in polar regions. The blank is just the problem of the tracer particle method and the region does not affect the nucleosynthesis outcome in the ejecta. Additionally, we have performed a convergence test of the nucleosynthesis results 
inside the jet by changing the number of distributed particles and confirmed that the results are not changed much by the 
difference in Paper I. 
Particles that appear deep inside the original iron core go through 
high-density and high-temperature regions; if the temperature is greater than 10$^{10}$ K, 
the compositions are determined under nuclear statistical equilibrium (NSE) condition: they are obtained 
from the values of ($\rho$,\,$T$,\,$Y_e$). 
Since these particles suffer from electron captures, we need to calculate the change 
in $Y_e$ of the ejected tracer particles due to the weak interactions 
of $e^{\pm}$ captures and $\beta^{\pm}$ decays until the last stage of NSE before the network calculation. 
The change in $Y_e$ is given by~\cite{ref:nishimura_2006}
\begin{equation}
\frac{d Y_e}{dt} = \sum_{i} (\lambda_{+} - \lambda_{-}) y_i,
\end{equation}
where $\lambda_{+}$ represents the $\beta^{-}$ and positron capture rates
 and $\lambda_{-}$ represents the $\beta^{+}$ and electron capture rates. 
Figure~\ref{fig:hist_ye} shows the distribution of the ejected mass in $M_{\odot}$ against the electron fraction 
of ejected particles at the end of NSE ($Y_{e, f}$). 
After the end of NSE, the nucleosynthesis calculation is done along the Lagrange evolution of each particle 
by using a large nuclear reaction network. 
Since the time of hydrodynamical simulation is insufficient to follow the decays of radioactive nuclei, 
the density and temperature profiles of particles are extrapolated assuming adiabatic expansion 
as in Refs.~\citen{ref:fujimoto_2007} and~\citen{ref:ono_2009}. We continue the nucleosynthesis calculation of the 
radioactive decays until $\sim$ 10$^{10}$ yr after the explosion. 
Note that we neglect the feedback of energy generations due to nuclear processes such as photodisintegrations 
in the hydrodynamical calculation because our nucleosynthesis calculations are just 
postprocessing. 
The effects of neutrino absorptions on the evolution of $Y_e$ should be noted here because we neglect the effects 
in the nucleosynthesis calculations, but it could be critical for the nucleosynthesis outcome. 
The inner edge of the accretion disk of the R51 model has $\rho \sim$ 10$^{9}$ g cm$^{-3}$ and the 
estimated neutrino luminosity ranges from $\sim$ 10$^{51}$ to $\sim$ 10$^{52}$ erg s$^{-1}$, 
which is about one order of magnitude smaller 
than that of canonical core-collapse supernovae. Therefore, the neutrino absorptions could not be effective. 
Note that the range of the neutrino luminosity does not differ much from that of other models in Paper I. 
Additionally, although the progenitor and specified initial angular momentum and magnetic field distributions 
differ from that of our model, recent nucleosynthesis calculations in a magnetorotationally driven core-collapse supernova model 
with an approximate neutrino transport scheme including effects of neutrino absorptions on $Y_e$ have 
revealed~\cite{ref:winteler_2012} 
that the peak distribution of $Y_e$ in the ejecta is shifted from $\sim$ 0.17 to $\sim$ 0.15 between with and without 
neutrino absorptions, but the results of the nucleosynthesis are not much affected by the difference. 
From the above considerations, the effects of neutrino absorptions on $Y_e$ in our model would be small and we 
neglect the effects. 


To investigate the heavy-element nucleosynthesis including the {\it p}- and {\it r}-processes, 
we calculate the nucleosynthesis along some Lagrange tracks of the ejected particles explained above 
using the large 
nuclear reaction network~\cite{ref:nishimura_2006} including 4463 nuclei (up to $^{292}$Am), 
where the reaction rates are mainly based on the REACLIB database~\cite{ref:rauscher_2000,ref:rauscher_2001} 
and the adopted theoretical mass formula is the extended Thomas-Fermi plus Strutinsky integral 
(ETFSI).\cite{ref:goriely_2001} 
It is noted that the reaction rates in the hydrostatic nucleosynthesis calculations are based on the JINA REACLIB database, 
which use the theoretical mass formula of the finite-range droplet model (FRDM).\cite{ref:meoller_1995} 
Therefore, there could be some inconsistency between the hydrostatic and explosive 
nucleosynthesis calculations. However, taking into account the marked uncertainty of the reaction rates far from 
the valley of the nuclear stability, the differences are acceptable in the present purpose. 

\subsection{{\it P}- and {\it r}-processes}

To investigate the total yield of the ejecta, we combine the results of the hydrostatic 
and explosive nucleosyntheses. 
We assume that all the unshocked matter located at larger radii than 
the jet front in the star (including the hydrogen envelope) of $\theta<$ 15$^{\circ}$ is 
successfully ejected by the jet, keeping the precollapse compositions unchanged. 

As shown in Fig.~\ref{fig:hist_ye}, the mass distribution of the ejecta tends to decrease 
as the electron fraction $Y_{e, f}$ decreases. 
The lowest value of $Y_{e, f}$ is 0.192 among the ejected particles. 
%
%
%
We find that the low-$Y_{e, f}$ particles are ejected from deep inside the disk, which 
are originally located from the edge of the iron core to the inner Si-rich layer and fall into near 
the BH due to gravitational collapse. 
The low-$Y_{e, f}$ particles strongly suffer from electron captures, 
which reduces the electron fractions of the particles. 
The distributions of compositions of the particles are initialized by the final state in NSE. 
On the other hand, the particles that do not suffer from nuclear burning are just 
pushed up by the inner jet at larger radii. Such particles maintain the precollapse compositions.  
Therefore, the precollapse abundances are crucial in part to determine the total compositions of the ejected matter. 

\begin{figure}
\begin{center}
\includegraphics[width=14cm,keepaspectratio,clip]{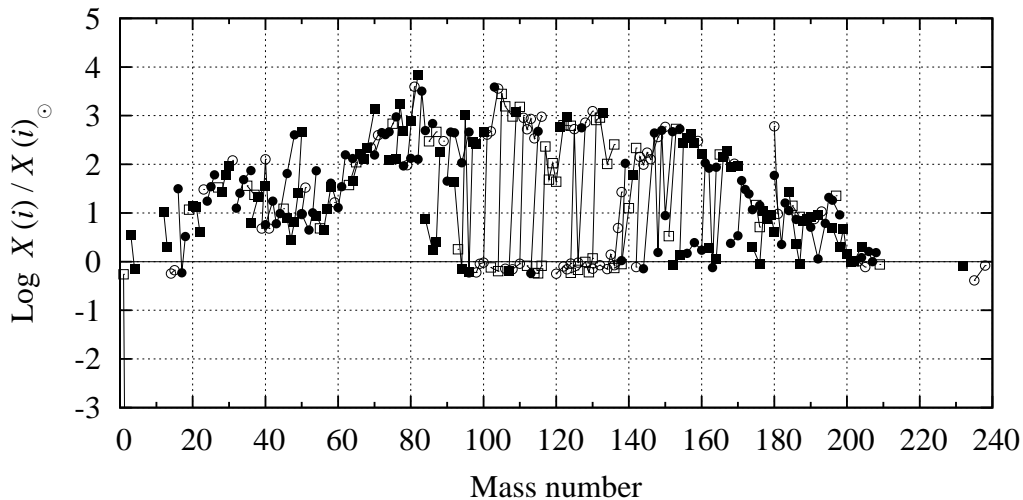}
\end{center}
\caption{Same as Fig.~\ref{fig:abund_over_631} but for the ejecta with the use of the larger network. 
}
\label{fig:abund_over}
\end{figure}

\begin{figure}
\begin{center}
\includegraphics[width=12cm,keepaspectratio,clip]{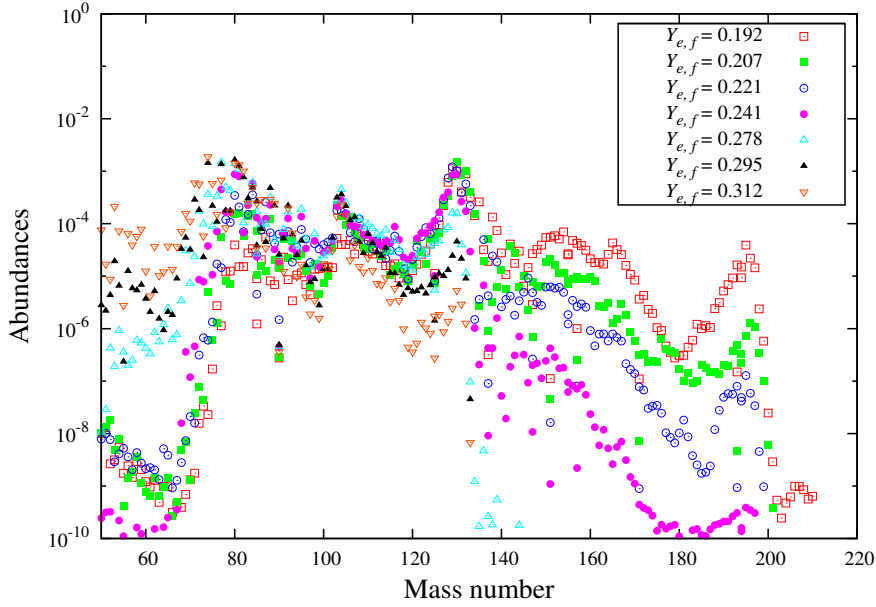}
\end{center}
\caption{Abundances of ejected particles with $Y_{e, f}$ that have different electron fractions at the end of NSE stage. 
}
\label{fig:abund_ye}
\end{figure}

The final overproduction factors $X(i)/X(i)_{\,\odot}$ averaged over the ejecta against the mass number are 
shown in Fig.~\ref{fig:abund_over}. Symbols connected by lines indicate the isotopes. 
Note that neutron-rich elements of $45<A<55$ and $60<A<160$ are highly overproduced relative to the solar values. 
Figure~\ref{fig:abund_ye} shows the abundances of ejected particles that have different electron fractions 
at the end of the NSE stage. The overproduced elements of $140<A<200$ 
originate from the ejected matter, which has a lower $Y_{e, f}$ of around 0.2 (Fig.~\ref{fig:abund_ye}), 
where the ejected materials undergo $r$-process nucleosynthesis. 
On the other hand, the ejected particles of $Y_{e, f}\sim$ 0.3 
produce elements of $60<A<90$. 
The overproduced elements of $A>90$ are primarily synthesized in the jet except for $^{180}$Ta.  
The overproduction factors have a peak at $A = 195$ (the neutron magic number of 126). 
It should be noted that our jet model cannot considerably produce the elements around the third peak. 
In contrast, 
they are significantly produced in the study by Fujimoto et al.~\cite{ref:fujimoto_2007,ref:fujimoto_2008}, 
which is attributed to the different 
distribution of the lowest part of $Y_{e, f}$ of ejecta. 
In Fujimoto et al., the particles with $Y_{e, f} \sim$ 0.1 are also ejected from their collapsar model of 40 $M_{\odot}$ 
(see e.g., Fig. 5 in Ref.~\citen{ref:fujimoto_2008}). 
Since the lowest $Y_{e, f}$ is about 0.2 in our model, strong {\it r}-process and fissions 
do not proceed. 
The difference in the distribution of $Y_{e, f}$ 
may be ascribed to those of the progenitors and implemented initial distributions of the angular 
momentum and the magnetic field. 
In particular, we include the initial toroidal magnetic field, which may inhibit matter to fall deep inside the core 
and suffer from strong electron captures. 

{\it S}-elements of $60<A<90$ are overproduced significantly, 
which is due to the weak {\it s}-process in the hydrostatic evolutional stage.
In contrast, no {\it s}-elements of $A>90$ are overproduced, 
which might be compensated by the products of the main {\it s}-process in the relatively low mass AGB stars. 

\begin{figure}
\begin{center}
\includegraphics[width=11cm,keepaspectratio,clip]{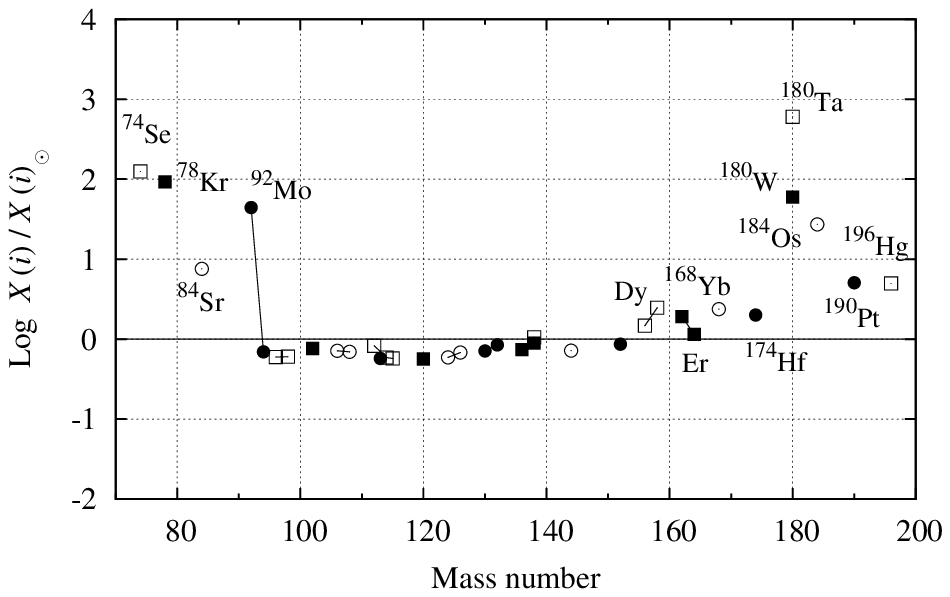}
\end{center}
\caption{Same as Fig.~\ref{fig:abund_over} but 35 {\it p}-elements.
}
\label{fig:p_nucl}
\end{figure}

\begin{figure}
\begin{center}
\includegraphics[width=10cm,keepaspectratio,clip]{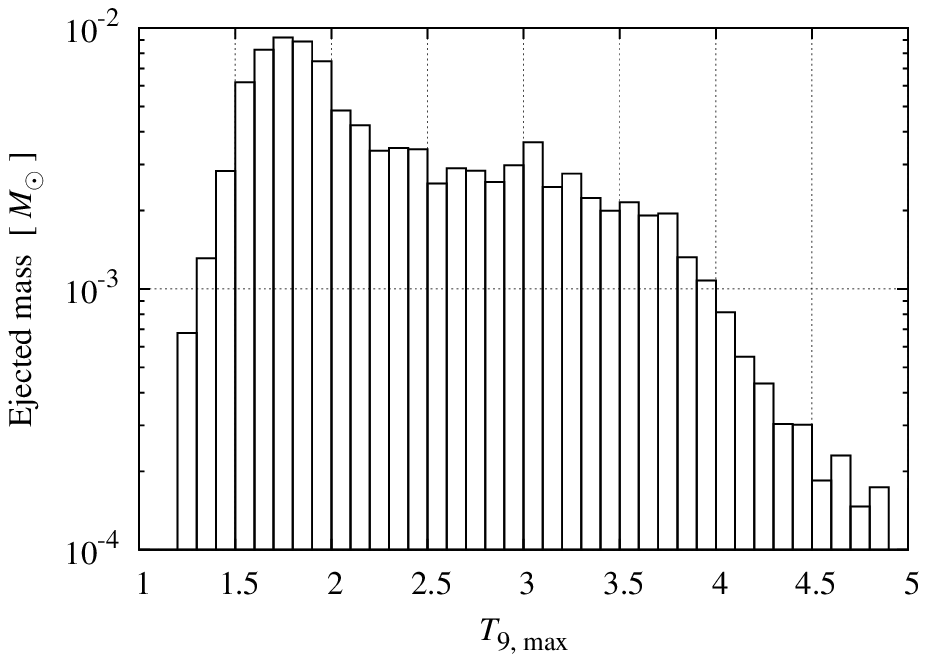}
\end{center}
\caption{Ejected mass originated from initial oxygen- and neon-rich layers against the peak temperature $T_{\rm 9,\,max}$ 
($T_{9} = T/\,(10^{9} {\rm K})$).
}
\label{fig:temp_hist}
\end{figure}

%
%
The overproduction factors of 35 {\it p}-elements are shown in Fig.~\ref{fig:p_nucl}. 
{\it P}-elements whose overproduction factors are greater than 10 are 
$^{74}$Se, $^{78}$Kr, $^{84}$Sr, $^{92}$Mo, $^{180}$Ta, $^{180}$W, $^{184}$Os, $^{190}$Pt, and $^{196}$Hg. 
On the other hand, {\it p}-elements of $^{74}$Se, $^{78}$Kr, $^{84}$Sr, $^{180}$W, and $^{180}$Ta are overproduced 
in the hydrostatic nucleosynthesis (see Fig.~\ref{fig:p_nucl_631}); 
those of $^{92}$Mo, $^{184}$Os, $^{190}$Pt, and $^{196}$Hg are mainly overproduced in explosive nucleosynthesis. 
Underproduced {\it p}-elements in the previous study of 
the {\it p}-process in Type II supernovae~\cite{ref:rayet_1995} such as $^{94}$Mo and $^{96,98}$Ru are not 
produced in our explosion model. 
Rayet et al.~\cite{ref:rayet_1990,ref:rayet_1995} have investigated the {\it p}-process (gamma process) 
in parametrized supernova explosion models. 
They concluded that the production of {\it p}-elements is very sensitive to the maximum temperature 
reached during the explosion, and the light ($N \lesssim 50$), intermediate ($50 \lesssim N \lesssim 82$)
and heavy ($N \gtrsim 82$) {\it p}-elements are produced in the peak temperature classified as 
$T_{\rm 9,\,max}\gtrsim3$, 
$3 \gtrsim T_{\rm 9,\,max} \gtrsim 2.7$, $2.5 \gtrsim T_{\rm 9,\,max}$, respectively, 
where $T_{9} = T/\,(10^{9} \,{\rm K})$. 
The mass distribution against the peak temperatures of the ejecta initially located in the oxygen- and neon-rich layers 
is shown in Fig.~\ref{fig:temp_hist}. The major part of the ejected matter has 
$T_{\rm 9,\,max} \lesssim 2.5$. 
Thereby, heavy {\it p}-elements are considerably produced in the explosive nucleosynthesis (Fig.~\ref{fig:p_nucl}). 
It is emphasized that light {\it p}-elements of $^{74}$Se, $^{78}$Kr, $^{84}$Sr, and $^{92}$Mo are primarily 
synthesized in the ejecta of $T_{\rm 9,\,max} \sim 16$ and $Y_{e, f} \sim 0.48$ initially located in 
Si-rich layers. Since the peak temperature is very high, protons, neutrons and alpha particles 
are significantly produced by photodisintegrations of heavy nuclei, and the light {\it p}-elements are produced by 
a sequence of neutron captures and subsequent proton captures as described in Fujimoto et al.~\cite{ref:fujimoto_2007}, 
the scenario of which was originally proposed by Howard et al.~\cite{ref:howard_1991}. 
Since light {\it p}-elements are already produced in the hydrostatic nucleosynthesis, the increments 
due to the primary process in the explosive nucleosynthesis are not prominent except for $^{92}$Mo. 
Note that in Fujimoto et al.~\cite{ref:fujimoto_2007}, intermediate {\it p}-elements such as 
$^{133}$In, $^{115}$Sn, and $^{138}$La are also produced by fission only in the ejecta of 
$Y_{e, f} \sim 0.1$. In our model, all the ejected matter has $Y_{e, f} \gtrsim 0.2$; 
therefore, fission reactions are not effective and intermediate {\it p}-elements cannot be produced. 


\section{Summary and discussion}

We have investigated the nucleosynthesis in a massive star of 32 $M_{\odot}$ helium core with solar metallicity 
during the stellar evolution and the jetlike supernova explosion. In this section,
we summarize the results and discuss the uncertainties of the production of elements and contribution to 
the chemical evolution of galaxies (\S 4.1). 
We also give additional discussion by comparing with observations in \S 4.2. 

Hydrostatic nucleosynthesis: 1) {\it S}-elements of $60<A<90$ are highly overproduced relative 
to the solar abundances, which is similar to the weak {\it s}-process scenario proposed in previous studies.
\cite{ref:prantzos_1987,ref:langer_1989,ref:rayet_2000} 
2) Although photodisintegrations of seed {\it s}-elements during oxygen burning produce 
{\it p}-elements, the produced elements are disintegrated in the later stages except for $^{180}$Ta. 
3) Three elements, Sr, Y, and Zr, are not much overproduced compared with the solar values except for $^{96}$Zr. 

Explosive nucleosynthesis: 4) Elements of $90<A<160$ are significantly overproduced relative to the 
solar values. 5) The overproduced elements of $140<A<200$ originate from the ejected matter 
with lower $Y_{e, f}$ around 0.2, which results in the {\it r}-process. 
6) The {\it p}-process produces mainly heavy {\it p}-elements ($N>50$) because the peak temperatures 
in the oxygen- and neon-rich layers are relatively low. 
7) Light {\it p}-elements are  produced as primary ones in the ejected matter, which has a high peak temperature. 
8) Compared with the previous study of {\it r}-process nucleosynthesis calculations in a collapsar model of 40 $M_{\odot}$ by 
Fujimoto et al.~\cite{ref:fujimoto_2007,ref:fujimoto_2008}, our jet model cannot considerably produce both the elements 
around the third peak of the solar {\it r}-elements and 
intermediate {\it p}-elements. This may be attributed to the differences in the progenitor 
and the specified initial angular momentum and magnetic field distributions. 

After all, our supernova explosion model results in neutron-rich elements of $70<A<140$ and
weak {\it s}-elements of $60<A<90$. 
The origin of other underproduced elements would be ascribed to different explosion mechanisms of supernovae. 

Here, we try to deduce the qualitative constraint on the event rate for our explosion model. 
Let us assume canonical supernova explosion to be spherical and/or neutrino-driven aspherical ones such as 
in Refs.~\citen{ref:rauscher_2002} and~\citen{ref:fujimoto_2011}. 
The overproduction levels for 
the elements of 
$60<A<160$ are 1 -- 2 orders of magnitude higher than that 
for those 
of $20<A<60$ 
compared with 
canonical ones. 
If we neglect the mass loss, the total ejected mass of our jetlike explosion model is around 2 $M_{\odot}$, 
which is about one order of magnitude less than that of 
canonical ones. 
Unless the event rate of our model is comparable to or one order of magnitude less than that of canonical ones, 
the elements of $60<A<160$ are too produced to explain the solar system abundance pattern. 
Therefore, 
the event rate 
of our model could 
be one order of magnitude less than that of 
canonical ones. 
The origin of other underproduced elements would be ascribed to different types of supernovae. 
However, the constraint speculated here is not strict. 

\subsection{Uncertainties of production of elements related to the chemical evolution of galaxies} 

We discuss the uncertainties concerning overproductions. 
To investigate the weak {\it s}-process in the hydrostatic nucleosynthesis, we adopt  $\beta$-decay rates of TY87. 
However, the table of the rates for $\rho$ and $T$ only covers the temperature and density ranges 
in the helium burning stage. 
As noted in \S 2.2, we find that productions of neutron-rich elements of $85<A<100$, in particular, Sr, Y, Z, and 
Mo, are very sensitive to the adopted $\beta$-decay rates. 
Therefore, it is urgent to construct a table of the $\beta$-decay rates, which 
covers all evolutional stages. 

In the explosive nucleosynthesis calculation, we use the theoretical reaction rates based on 
ETFSI mass formula for elements far from the valley of the nuclear stability, 
in which  no experimental cross sections are available. 
The produced abundance pattern should depend on an adopted mass formula.\cite{ref:fujimoto_2007} 

We assume that all the matter that has larger radii 
than the jet front in the star of $\theta<$ 15$^{\circ}$ is successfully ejected by the jet. 
However, the angle from which the matter should be ejected is rather uncertain. 
Although we neglect the effects of the mass loss, a star of 70 $M_{\odot}$ with solar metallicity should suffer from 
a significant mass loss.\cite{ref:heger_2003_massive_star} If the mass loss is effective, 
almost all the hydrogen-rich envelope should be ejected as the stellar wind. 
If the above uncertainties are taken into account, the overproduction levels could become one order of magnitude less than 
those described in the previous sections. However, the overall abundance patterns do not change qualitatively 
except for the overproduction levels. 

\begin{figure}
\begin{center}
\includegraphics[width=13cm,keepaspectratio,clip]{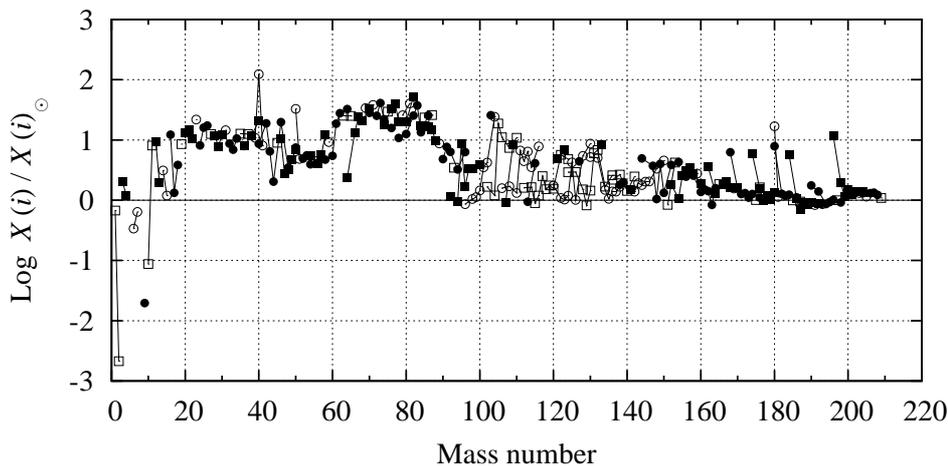}
\caption{Same as Fig.~\ref{fig:abund_over_631} but for averaged over progenitor masses weighed on the basis of Salpeter's IMF 
with the use of spherical explosion models from Rauscher et al. for 
15, 19, 20, 21, 25, 30, 35, and 40 $M_{\odot}$ progenitors 
and our explosion model for 70 $M_{\odot}$.
}
\label{fig:imf_average}
\end{center}
\end{figure}

\begin{figure}
\begin{center}
\includegraphics[width=13cm,keepaspectratio,clip]{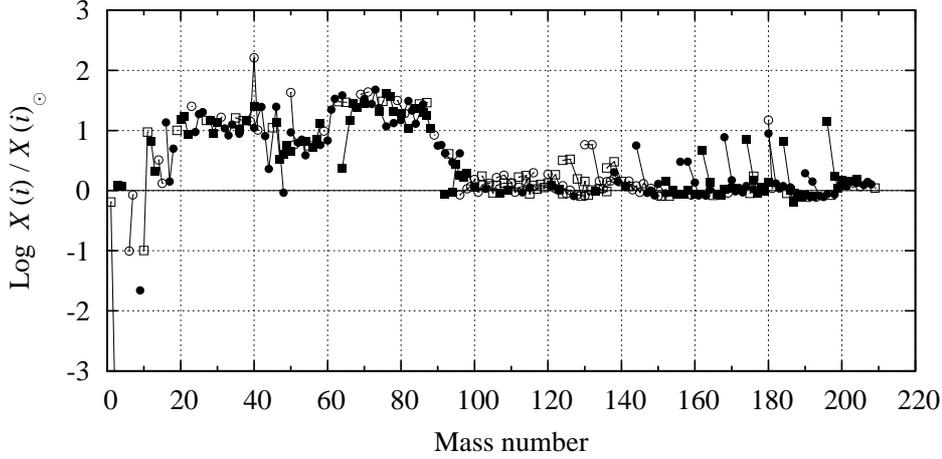}
\caption{
Same as Fig.~\ref{fig:imf_average} but for only from 15 to 40 $M_{\odot}$.
}
\label{fig:imf_average_without_70Msolar}
\end{center}
\end{figure}

To investigate the possible contribution of our explosion model to the chemical evolution of galaxies, 
we estimate overproduction factors averaged over progenitor masses weighted on the basis of 
Salpeter's stellar initial mass function (IMF).\cite{ref:salpeter_1955} 
The IMF averaged overproduction factor of the element $i$, $f_{\rm IMF}\,(i)$, is defined as 
\begin{equation}
f_{\rm IMF}\,(i) \equiv 
\frac{\displaystyle \int_{M_{1}}^{M_{2}} X\,(i,M)\, f_{\rm ej}(M)\, M \,\phi\,(M)\, dM \,/ \,X(i)_{\,\odot}}
{\displaystyle \sum_{j} \int_{M_{1}}^{M_{2}} X\,(j,M)\, f_{\rm ej}(M)\, M \, \phi\,(M)\, dM }, 
\label{eq:over_imf}
\end{equation}
where $X(i,M)$ is the mass fraction of the element $i$ in the ejecta, 
$M$ the initial progenitor mass, $f_{\rm ej}$ the mass 
ratio of the ejecta to the initial mass and 
$\phi\,(M) \propto M^{-2.35}$ the number of stars within the mass range between $M$ and $M+dM$, 
which is Salpeter's IMF. 
Let us adopt spherical postexplosion models from Rauscher et al.~\cite{ref:rauscher_2002} for progenitors of 
15, 19, 20, 21, 25, 30, 35, and 40 $M_{\odot}$\footnote{We take the nucleosynthesis data of postsupernova models from the web site: 
\url{http://homepages.spa.umn.edu/~alex/nucleosynthesis/RHHW02.shtml}. 
Note that the data of 30, 35, and 40 $M_{\odot}$ are not yet published, 
and necessary data for the integration in Eq. (\ref{eq:over_imf}) are obtained by interpolation.} 
with the solar metallicity and take our model for 70 $M_{\odot}$. 
During the integrations in equation (\ref{eq:over_imf}), necessary values of $X(i,M)$ and $f_{\rm ej}$ 
are obtained by interpolation of sample values. 
We take $M_{1}$ to be 15 $M_{\odot}$ and $M_{2}$ to be 70 $M_{\odot}$. 
For our 70 $M_{\odot}$ model, we assume all of the hydrogen envelope to be ejected as the stellar wind, because 
70 $M_{\odot}$ with the solar metallicity may strongly suffer from the mass loss.\cite{ref:heger_2003_massive_star} 
Figure \ref{fig:imf_average} shows the IMF averaged overproduction factors. 
We also show the overproduction factors averaged only from 15 $M_{\odot}$ to 40 $M_{\odot}$ in 
Fig.~\ref{fig:imf_average_without_70Msolar} for reference. 
We can see relatively high overproduction of $^{40}$K relative to the solar value, which arises from the large enhancement 
of the star of 20 $M_{\odot}$ (see Fig.~4 in Ref.~\citen{ref:rauscher_2002}). 
We also recognize relatively larger overproduction factors for elements of $60<A<90$. 
The overproduction level of $60<A<90$ is slightly enhanced by a few factors due to 
the inclusion of our 70 $M_{\odot}$ model 
(Fig. \ref{fig:imf_average}) 
compared with that averaged over the range only between 
15 $M_{\odot}$ and 40 $M_{\odot}$ 
(Fig. \ref{fig:imf_average_without_70Msolar}). 
Recall that 
the production of elements of $60<A<90$ results from the weak {\it s}-process. Overproduced neutron-rich elements of $90<A<160$ 
are attributed to the 70 $M_{\odot}$ model in which the elements are primarily synthesized in the explosive nucleosynthesis by 
the {\it r}-process. {\it P}-elements of $110<A<200$ are overproduced to the solar values, which mainly come from 
the {\it p}-process in 15 -- 40 $M_{\odot}$ stars. $^{180}$Ta is highly overproduced in our 70 $M_{\odot}$ model. 
Therefore, it is interesting whether our model could contribute to the solar $^{180}$Ta abundance. 
If we average the overproduction factors over the 
mass range only between 15 $M_{\odot}$ and 40 $M_{\odot}$ 
(Fig. \ref{fig:imf_average_without_70Msolar}), 
the IMF averaged overproduction factor of $^{180}$Ta is 1.18 
in logarithmic scale. On the other hand, if we include the 70 $M_{\odot}$ model in the integration in Eq. (\ref{eq:over_imf}), the 
overproduction factor becomes 1.23, 
which corresponds to a 12\% increase. 
Therefore, the contribution of 40 -- 70 $M_{\odot}$ to the solar $^{180}$Ta abundance is negligible in view of 
uncertainties of the present study. 
Overall, our model contributes to the solar weak {\it s}-elements of $70<A<90$ and 
neutron-rich elements of $90<A<160$. 
However, we should treat the results with caution because our 
jetlike explosion model is only one specific set of parameters of angular momentum and magnetic field distributions, and 
the fraction of such an aspherical explosion is highly uncertain. 
As suggested by Fujimoto et al.,~\cite{ref:fujimoto_2008} the {\it r}-process does not occur in a less energetic jet 
and the jet properties depend on specified parameters of the initial angular momentum and magnetic field distributions. 
If the fraction of jetlike explosions among the progenitors is less than 10\%, the contribution of jet-induced 
nucleosynthesis above 40 $M_{\odot}$ to the chemical evolution of galaxies would be minor. 
Therefore, we should regard the contributions of our model deduced here 
as the upper limits. 
In addition, we investigate only the progenitors with solar metallicity. 
As a consequence, some simulations of the chemical evolution of galaxies are required to ascertain the contribution especially 
for elements whose productions depend on the metallicity, which is beyond the scope of this paper. 

\subsection{Comparison with abundances in metal-poor stars and the possibility for LEPP}

\begin{figure}
\begin{center}
\includegraphics[width=11cm,keepaspectratio,clip]{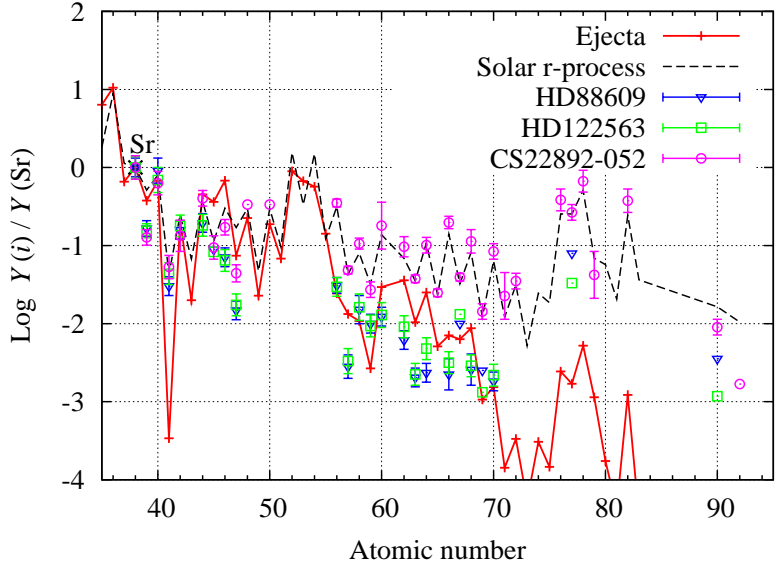}
\end{center}
\caption{Abundance ratios relative to Sr: calculated ejecta (solid line), 
solar system {\it r}-process elements (dashed line) (Simmerer et al.~\cite{ref:simmerer_2004}) 
and three very metal-poor stars.\cite{ref:honda_2007,ref:sneden_2003} 
Symbols with error bars indicate the observations of their abundances. 
Triangles, squares and circles represent HD88609, HD122563, and CS22892-052, respectively. 
Symbols without error bar indicate the upper limits.
}
\label{fig:r_element}
\end{figure}

Although our progenitor is assumed to have the solar metallicity, the production of {\it r}-elements, 
which are primarily synthesized without seeds, does not depend on the metallicity. 
Therefore, it is worthwhile to compare our results with 
the abundances of extremely metal-poor stars, which are not affected seriously by the {\it s}-process. 
It is noted that the overproduced elements of $Z>38$ are primarily produced in the explosive nucleosynthesis. 
Abundance ratios relative to Sr against the atomic number are shown in Fig.~\ref{fig:r_element}. 
Solid and dashed lines denote abundance ratios of our model and that of solar system 
{\it r}-process elements,~\cite{ref:simmerer_2004} respectively. Circles, triangles, and squares indicate 
the values of observations of very metal-poor stars CS22892-052~\cite{ref:sneden_2003}, 
HD88609~\cite{ref:honda_2007} and HD122563~\cite{ref:honda_2007}, respectively. 
The symbols without error bars indicate the upper limits. 
We can see that the abundance pattern of CS22892-052, which is a typical {\it r}-process-rich 
([Eu/Fe] $\gtrsim1$) star, coincides well with that of the solar {\it r}-element pattern, 
although some exceptions are recognized. 
The {\it r}-process-poor stars ([Eu/Fe] $\lesssim1$), HD88609 and HD122563, have a clearly 
decreasing trend as the atomic number increases. 
It is noted that although [Eu/Fe] values of {\it r}-process-poor stars are low compared with those of the 
{\it r}-process-rich ones, abundances in HD88609 ([Fe/H] $\sim -3.0$) and HD122563 ([Fe/H] $\sim -2.7$) should come 
from the weak {\it r}-process because the sources of the {\it s}-process, AGB stars, have not had sufficient 
time to evolve before the formation of such metal-poor halo stars~\cite{ref:honda_2007}. 
The abundances of the ejecta of our model show a decreasing trend in proportion to the decrease 
in the atomic number, which is similar to the case of HD88609 and HD122563.\cite{ref:roederer_2010} 
This result can be attributed to the decrease in the ejected masses at lower values of 
$Y_{e, f}$ (Fig.~\ref{fig:hist_ye}). 

The abundance ratios relative to the solar values are summarized in Table~\ref{table:abund_ratio}, where 
the observational values are taken from Sneden et al.~\cite{ref:sneden_2003} for CS22892-052 
and Honda et al.~\cite{ref:honda_2007} for HD88609 and HD122563. 
The values of [Sr/Fe], [Y/Fe] and [Zr/Fe] are similarly observed and the ejecta also has the same 
tendency. 
While [Sr/Eu] of the {\it r}-process-rich star CS22892-052 ([Sr/Eu] $\sim -1$) is very small compared 
with that of the {\it r}-process-poor stars HD88609 and HD122563 ([Sr/Eu] $\sim +0.3$), 
the value of the ejecta ([Sr/Eu] $\sim -0.4$) is closer to 
that of the {\it r}-process-poor stars than to that of the {\it r}-process-rich stars, 
which reflects the declining trend of abundances as the atomic number 
increases (Fig.~\ref{fig:r_element}). 

We find that Sr-Y-Zr isotopes are primarily synthesized in the explosive nucleosynthesis 
by a similar process of primary synthesis of light {\it p}-elements as described in \S3.3. 
The ejected matter of $T_{\rm 9,\,max} \sim 16$ and $Y_{e, f} \sim 0.45$ produces most
isotopes of Sr-Y-Zr, 
which is more neutron rich than that in the case of primary light {\it p}-element synthesis. 
In such high peak temperature and density, there exist protons, neutrons, and alpha particles, where 
a lot of neutrons are produced by electron captures. After the temperature decreases to 8$\times$10$^{9}$ K, 
a sequence of neutron captures and $\beta^{-}$-decays 
produces slightly neutron-rich Sr-Y-Zr isotopes from lighter elements. After the exhaustion of neutrons, 
proton captures and gamma processes follow. Recall that $^{96}$Zr is overproduced relative to the solar value 
in the hydrostatic nucleosynthesis (Fig.~\ref{fig:abund_over_631}). 
$^{88}$Sr, $^{89}$Y, and $^{91,\,92,\,94,\,96}$Zr are highly overproduced relative to the solar values due to 
the primary process in the explosive nucleosynthesis. It is noted that 
$^{96}$Zr is more produced in the explosive nucleosynthesis. 

Travaglio et al.~\cite{ref:travaglio_2004} have suggested that based on a galactic chemical evolution (GCE) model, 
a primary process from massive stars (LEPP) other than the general {\it s}- and {\it r}-processes 
is needed to explain 8\% of the solar abundance for Sr and 18\% of the solar Y and Zr abundances. 
In their GCE model, the yields of the {\it s}-process have been derived from AGB models 
and they have also added a small contribution ($\sim10$\% of the solar ones) from the weak {\it s}-component for Sr. 
It is emphasized that the contribution from the {\it r}-process has been deduced 
from the very {\it r}-process-rich CS22892-052~\cite{ref:sneden_2003}, that is, contributions to {\it r}-elements 
from {\it r}-process-{\it poor} stars like our explosion model have not been included. 
In our explosion model, the ejecta has a larger [Sr/Eu] than that of {\it r}-process-rich stars, and 
Sr, Y, and Zr are mainly produced by the primary process. Therefore, our explosion model could 
be one of the sites of LEPP. However, the calculations are limited to only one model of the progenitor with the solar metallicity 
and specific set of parameters of initial distribution of magnetic field and angular momentum. 
As suggested in Ref. \citen{ref:fujimoto_2008}, ejected masses of {\it r}-elements depend on the jet 
properties such as the explosion energies. 
Therefore, the effects of explosion models on the chemical evolution of galaxies remain uncertain
and should be studied in the future. 



\begin{table}
\begin{center}
\caption{Abundance ratios relative to the solar values for extremely metal-poor stars and our model 
indicated by ``Ejecta''.}
\label{table:abund_ratio}
\renewcommand{\arraystretch}{1.3}
\begingroup
\small
\begin{tabular}{lcccccccc}
\vspace{-0.2cm} \\
\hline\hline &
{[Fe/H]} &
{[Sr/Fe]} &
{[Y/Fe]} &
{[Zr/Fe]} &
{[Eu/Fe]} &
{[Sr/Eu]} 
\vspace{+0.1cm} 
 \\ 
\hline
CS22892-052  &  $-3.1$  & $+0.6$  & $+0.44$ & $+0.78$ & $+1.64$ & $-1.04$ \\
HD88609 &  $-3.0$  & $-0.05$  & $-0.12$ & $+0.24$ &  $-0.33$ & $+0.28$ \\
HD122563  & $-2.7$  & $-0.27$  & $-0.37$ & $-0.10$ & $-0.52$ & $+0.25$ \\
Ejecta &  --  & $+1.50$  & $+1.79$ & $+1.65$ & $+1.90$ & $-0.4$ \\
\hline
\end{tabular}
\endgroup
\end{center}
\end{table}

\section*{Acknowledgements}
We would like to thank A. Heger and his collaborators for offering their data. 
M. Ono thanks N. Nishimura for stimulating discussion. 
K. Kotake is grateful to K. Sato for continuous encouragement. 
This work has been supported in part by Grants-in-Aid for Scientific Research 
(Nos.~18540279,~19104006,~20740150,~22540297 and 
24540278) 
from the Ministry of Education, Culture, Sports, Science and Technology of Japan. 

\appendix
\section{Thermal Population of Ground and Isomeric States of $^{180}{\rm Ta}$}

$^{180}$Ta is one of the rarest isotopes in the solar system and it has a long-lived isomeric state. 
The isomeric state has $J^{\pi} = 9^{-}$ and the half-life $\tau_{1/2}$ is 1.2$\times10^{15}$ yr, 
while the ground state has $J^{\pi} = 1^{+}$ and $\tau_{1/2} \simeq 8.152$ h. 
We treat specially the reaction rates concerning $^{180}$Ta by a method similar to that described in Ref.~\citen{ref:rauscher_2002}. 

Because of the selection rule for the spin and parity, the isomeric state ($^{180{\rm m}}$Ta) cannot directly decay 
into the grand state ($^{180{\rm g}}$Ta). 
However, if the temperature is sufficiently high, $^{180{\rm m}}$Ta can decay into the ground state through 
thermally excited states. 
In stellar interiors during some burning stages and in supernova explosions, the two states of $^{180}$Ta 
are thermally populated. 
In thermal equilibrium, the population ratio $P_{\rm iso}$ of the isomer relative to the ground state 
is given as~\cite{ref:rauscher_2002}
\begin{equation}
P_{\rm iso} = \frac{(2J_{\rm iso} + 1) \exp (-E_{\rm iso}/kT)}{(2J_{\rm gs} + 1)} 
= \frac{19}{3} e^{-0.8738/T_9},
\label{eq:pop_iso}
\end{equation}
where $J_{\rm gs}$ and $J_{\rm iso}$ are the spins of the ground and isomeric states, respectively, and 
$E_{\rm iso}$ the excitation energy of the isomer, $T_{9} = T/10^{9}$ K. 
If the temperature decreases to a critical temperature $T_{\rm crit}$, the two states 
no longer interact with each other. 
In the explosive burning scenario, the critical temperature is crucial to determine the amount of 
$^{180{\rm m}}$Ta.  
Belic et al.~\cite{ref:belic_1999} have derived the effective decay rate of $^{180{\rm m}}$Ta by 
photoactivation experiments. 
We simply assume $T_{\rm crit}$ as 0.35$\times$10$^{9}$ K from the temperature-dependent 
decay rate (Fig. 4 in Ref.~\citen{ref:belic_1999}). 
The effective $^{180}$Ta rates of neutron-induced reaction and $\beta$-decays are given by
\begin{equation}
\lambda_{\rm eff} = f_{\rm gs} \lambda_{\rm gs} + f_{\rm iso} \lambda_{\rm iso}
\label{eq:lambda_eff},
\end{equation}
where $f_{\rm gs}$ and $f_{\rm iso}$ are the fractions of the ground and isomeric states, respectively. 
For this effective decay rate, we derive $\lambda_{\rm gs}$ from the half-life of the ground state and 
$\lambda_{\rm iso}$ from Belic et al.~\cite{ref:belic_1999} For the neutron capture of $^{180}$Ta, 
$\lambda_{\rm gs}$ is taken from Ref.~\citen{ref:rauscher_2002} and $\lambda_{\rm iso}$ is 
taken from the JINA REACLIB database.\cite{ref:cyburt_2010}
If $T<T_{\rm crit}$, we make all the $^{180{\rm g}}$Ta decayed by hand and set $f_{\rm gs}$ to be 0.

%


\begin{thebibliography}{99}

\bibitem{ref:fujimoto_2007} 
    	S.~Fujimoto, M.~Hashimoto, K.~Kotake and S.~Yamada, \AJ{656,2007,382}.
	
\bibitem{ref:fujimoto_2008}
		S.~Fujimoto, N.~Nishimura and M.~Hashimoto, \AJ{680,2008,1350}. 

\bibitem{ref:thielemann_2011} 
	F.-K.~Thielemann, A.~Arcones, R.~K{\"a}ppeli, M.~Liebend{\"o}rfer, T.~Rauscher, C.~Winteler, 
	C.~Fr{\"o}hlich, I.~Dillmann, T.~Fischer, G.~Martinez-Pinedo, K.~Langanke, K.~Farouqi, 
	K.-L.~Kratz, I.~Panov and I.~K.~Korneev, \JL{Prog.\ Part.\ Nucl.\ Phys.,66,2011,346}.


\bibitem{ref:bbfh_1957}
	E.~M.~Burbidge, G.~R.~Burbidge, W.~A.~Fowler and F.~Hoyle, \JL{Rev.\ Mod.\ Phys.,29,1957,547}.


\bibitem{ref:hoffman_1997} 
	R.~D.~Hoffman, S.~E.~Woosley and Y.-Z.~Qian, \AJ{482,1997,951}.

\bibitem{ref:otsuki_2000}
	K.~Otsuki, H.~Tagoshi, T.~Kajino and S.~Wanajo, \AJ{533,2000,424}.


\bibitem{ref:wanajo_2007}
	S.~Wanajo, \AJ{666,2007,L77}.

\bibitem{ref:kuroda_2008}
	T.~Kuroda, S.~Wanajo and K.~Nomoto, \AJ{672,2008,1068}.

\bibitem{ref:fischer_2010} 
	T.~Fischer, S.~C.~Whitehouse, A.~Mezzacappa, F.-K.~Thielemann and M.~Liebend{\"o}rfer, 
	\JL{Astron.\ Astrophys.,517,2010,A80}.
	

\bibitem{ref:metzger_2010}
	B.~D.~Metzger, G.~Mart{\'{\i}}nez-Pinedo, S.~Darbha, E.~Quataert, A.~Arcones, D.~Kasen, 
	R.~Thomas, P.~Nugent, I.~V.~Panov and N.~T.~Zinner, 
	\JL{Mon.\ Not.\ R.\ Astron.\ Soc.,406,2010,2650}.


\bibitem{ref:roberts_2011}
	L.~F.~Roberts, D.~Kasen, W.~H.~Lee and E.~Ramirez-Ruiz, \AJ{736,2011,L21}.

\bibitem{ref:goriely_2011}
	S.~Goriely, A.~Bauswein and H.-T.~Janka, \AJ{738,2011,L32}.

\bibitem{ref:wanajo_2011}
	S.~Wanajo and H.-T.~Janka, \AJ{746,2012,180}.

\bibitem{ref:kappeler_2011}
	F.~K\"{a}ppeler, R.~Gallino, S.~Bisterzo and W.~Aoki, \JL{Rev.\ Mod.\ Phys.,83,2011,157}.

\bibitem{ref:rayet_1995} 
	M.~Rayet, M.~Arnould, M.~Hashimoto, N.~Prantzos and K.~Nomoto, \JL{Astron.\ Astrophys.,298,1995,517}.

\bibitem{ref:roederer_2010}
	I.~U.~Roederer, J.~J.~Cowan, A.~I.~Karakas, K.-L.~Kratz, M.~Lugaro, J.~Simmerer, K.~Farouqi and C.~Sneden, 
	\AJ{724,2010,975}.
	


\bibitem{ref:sneden_2003} 
	C.~Sneden, J.~J.~Cowan, J.~E.~Lawler, I.~I.~Ivans, S.~Burles, T.~C.~Beers, F.~Primas, V.~Hill, 
	J.~W.~Truran, G.~M.~Fuller, B.~Pfeiffer and K.-L.~Kratz, \AJ{591,2003,936}.

\bibitem{ref:travaglio_2004}
	C.~Travaglio, R.~Gallino, E.~Arnone, J.~Cowan, F.~Jordan and C.~Sneden, \AJ{601,2004,864}.
	
\bibitem{ref:wang_2002}
		L.~Wang, J.~C.~Wheeler, P.~H{\"o}flich, A.~Khokhlov, D.~Baade, D.~Branch, P.~Challis, A.~V.~Filippenko, 
		C.~Fransson, P.~Garnavich, R.~P.~Kirshner, P.~Lundqvist, R.~McCray, N.~Panagia, C.~S.~J.~Pun, 
		M.~M.~Phillips, G.~Sonneborn and N.~B.~Suntzeff, \AJ{579,2002,671}.
		

\bibitem{ref:tanaka_2007}
		M.~Tanaka, K.~Maeda, P.~A.~Mazzali and K.~Nomoto, \AJ{668,2007,L19}.

\bibitem{ref:woosley_2005}
		S.~Woosley and T.~Janka, \JL{Nat.\ Phys.,1,2005,147}.

\bibitem{ref:kotake_2006_rev}
		K.~Kotake, K.~Sato and K.~Takahashi, \JL{Rep.\ Prog.\ Phys.,69,2006,971}.	

\bibitem{ref:kotake_2011_rev}
		K.~Kotake, arXiv:1110.5107.

\bibitem{ref:kotake_2012_rev1}
		K.~Kotake, T.~Takiwaki, Y.~Suwa, W.~Iwakami Nakano, S.~Kawagoe, Y.~Masada and S.~Fujimoto, arXiv:1204.2330.
		
\bibitem{ref:kotake_2012_rev2}
		K.~Kotake, K.~Sumiyoshi, S.~Yamada, T.~Takiwaki, T.~Kuroda, Y.~Suwa and H.~Nagakura, arXiv:1205.6284.




\bibitem{ref:marek_2009}
		A.~Marek and H.-T.~Janka, \AJ{694,2009,664}.
		
\bibitem{ref:suwa_2010} 
		Y.~Suwa, K.~Kotake, T.~Takiwaki, S.~C.~Whitehouse, M.~Liebend{\"o}rfer and K.~Sato, 
		\JL{Publ. Astron. Soc. Jpn.,62,2010,L49}.

\bibitem{ref:takiwaki_2011}
		T.~Takiwaki, K.~Kotake and Y.~Suwa, arXiv:1108.3989.

\bibitem{ref:kuroda_2012}
		T.~Kuroda, K.~Kotake and T.~Takiwaki, arXiv:1202.2487.


\bibitem{ref:fischer_2011} 
	T.~Fischer, I.~Sagert, G.~Pagliara, M.~Hempel, J.~Schaffner-Bielich, T.~Rauscher, 
	F.-K.~Thielemann, R.~K{\"a}ppeli, G.~Mart{\'{\i}}nez-Pinedo and M.~Liebend{\"o}rfer, 
	\JL{Astrophys.\ J.\ Suppl.,194,2011,39}.

\bibitem{ref:kotake_2004}
        K.~Kotake, H.~Sawai, S.~Yamada and K.~Sato, \AJ{608,2004,391}.        

\bibitem{ref:sawai_2005}
		H.~Sawai, K.~Kotake and S.~Yamada, \AJ{631,2005,446}.

\bibitem{ref:shibata_2006}
		M.~Shibata, Y.~T.~Liu, S.~L.~Shapiro and B.~C.~Stephens, \PRD{74,2006,104026}.

\bibitem{ref:suwa_2007}
		Y.~Suwa, T.~Takiwaki, K.~Kotake and K.~Sato, 
		\JL{Publ.\ Astron.\ Soc.\ Jpn.,59,2007,771}.
		
\bibitem{ref:takiwaki_2009}
		T.~Takiwaki, K.~Kotake and K.~Sato, \AJ{691,2009,1360}.

\bibitem{ref:heger_2003_massive_star}
        A.~Heger, C.~L.~Fryer, S.~E.~Woosley, N.~Langer and D.~H.~Hartmann, \AJ{591,2003,288}.

\bibitem{ref:macfadyen_woosley_1999}
		A.~I.~MacFadyen and S.~E.~Woosley, \AJ{524,1999,262}.
%
\bibitem{ref:harikae_2010}
		S.~Harikae, K.~Kotake, T.~Takiwaki and Y.~Sekiguchi, \AJ{720,2010,614}.

\bibitem{ref:Zalamea_2011}	    
	    I.~Zalamea and A.~M.~Beloborodov, \JL{Mon.\ Not.\ R.\ Astron.\ Soc.,410,2011,2302}.

\bibitem{ref:woosley_1993}
	    S.~E.~Woosley, \AJ{405,1993,273}.

%

\bibitem{ref:koide_2002}
		S.~Koide, K.~Shibata, T.~Kudoh and D.~L.~Meier, \JL{Science,295,2002,1688}.

\bibitem{ref:proga_2003}
		D.~Proga, A.~I.~MacFadyen, P.~J.~Armitage and M.~C.~Begelman, \AJ{599,2003,L5}.

\bibitem{ref:mizuno_2004}
		Y.~Mizuno, S.~Yamada, S.~Koide and K.~Shibata, \AJ{615,2004,389}.

\bibitem{ref:fujimoto_2006} 
	S.~Fujimoto, K.~Kotake, S.~Yamada, M.~Hashimoto and K.~Sato, \AJ{644,2006,1040}.


\bibitem{ref:nagataki_2007}
		S.~Nagataki, R.~Takahashi, A.~Mizuta and T.~Takiwaki, \AJ{659,2007,512}.

\bibitem{ref:harikae_2009}
		S.~Harikae, T.~Takiwaki and K.~Kotake, \AJ{704,2009,354}.

\bibitem{ref:nagataki_2009}
		S.~Nagataki, \AJ{704,2009,937}.

\bibitem{ref:nagataki_2011}
		S.~Nagataki, \JL{Publ. Astron. Soc. Jpn.,163,2011,1243}.
		
\bibitem{ref:blandford_1977}
		R.~D.~Blandford and R.~Znajek, \JL{Mon.\ Not.\ R.\ Astron.\ Soc.,179,1977,433}.

	


\bibitem{ref:nagataki_2003}
		S.~Nagataki, A.~Mizuta, S.~Yamada, H.~Takabe and K.~Sato, \AJ{596,2003,401}.

\bibitem{ref:nagataki_2006}
		S.~Nagataki, A.~Mizuta and K.~Sato, \AJ{647,2006,1255}.
		
\bibitem{ref:winteler_2012}
		C.~Winteler, R.~K{\"a}ppeli, A.~Perego, A.~Arcones, N.~Vasset, N.~Nishimura, M.~Liebend{\"o}rfer and F.-K.~Thielemann, 
		\AJ{750,2012,L22}
		.
		
\bibitem{ref:rauscher_2002} 
	T.~Rauscher, A.~Heger, R.~D.~Hoffman and S.~E.~Woosley, \AJ{576,2002,323}.	

\bibitem{ref:woosley_1995} 
	S.~E.~Woosley and T.~A.~Weaver, \JL{Astrophys.\ J.\ Suppl.,101,1995,181}.


\bibitem{ref:fujimoto_2011}
	S.~Fujimoto, K.~Kotake, M.~Hashimoto, M.~Ono and N.~Ohnishi, \AJ{738,2011,61}.

\bibitem{ref:ono_2009}  
	M.~Ono, M.~Hashimoto, S.~Fujimoto, K.~Kotake and S.~Yamada, \PTP{122,2009,755}.


\bibitem{ref:hashimoto_1995} 
	M.~Hashimoto, \PTP{94,1995,663}.

\bibitem{ref:puls_2008}
		J. Puls, J. S. Vink and F. Najarro, \JL{Astron.\ Astrophys.\ Rev.,16,2008,209}.

\bibitem{ref:prantzos_1987} 
	N.~Prantzos, M.~Arnould and J.-P.~Arcoragi, \AJ{315,1987,209}.

\bibitem{ref:cyburt_2010} 
	R.~H.~Cyburt, A.~M.~Amthor, R.~Ferguson, Z.~Meisel, K.~Smith, S.~Warren, A.~Heger, R.~D.~Hoffman, T.~Rauscher, 
	A.~Sakharuk, H.~Schatz, F.~K.~Thielemann and  M.~Wiescher, \JL{Astrophys.\ J.\ Suppl.,189,2010,240}.

\bibitem{ref:dillmann_2006}
	I.~Dillmann, M.~Heil, F.~K{\"a}ppeler, R.~Plag, T.~Rauscher and F.-K.~Thielemann, 
	\JL{{AIP Conf. Ser., Capture Gamma-Ray Spectroscopy and Related Topics, ed. A.~Woehr \& A.~Aprahamian},819,2006,123}

\bibitem{ref:takahashi_1987} 
	K.~Takahashi and K.~Yokoi, \JL{At.\ Data\ Nucl. Data\ Tables,36,1987,375}.

\bibitem{ref:anders_1989}
	E.~Anders and N.~Grevesse, \JL{Geochim.\ Cosmochim.\ Acta,53,1989,197}.	

\bibitem{ref:langer_1989} 
	N.~Langer, J.-P.~Arcoragi and M.~Arnould, \JL{Astron.\ Astrophys.,210,1989,187}.


\bibitem{ref:rayet_2000} 
	M.~Rayet and M.~Hashimoto, \JL{Astron.\ Astrophys.,354,2000,740}.


\bibitem{ref:arnould_1976}
		M.~Arnould, \JL{Astron.\ Astrophys.,46,1976,117}.
		
\bibitem{ref:nomoto_1988}
		K.~Nomoto and M.~Hashimoto, \PRP{163,1988,13}.

\bibitem{ref:hashimoto_1989}
		M.~Hashimoto, K.~Nomoto and T.~Shigeyama, \JL{Astron.\ Astrophys.,210,1989,L5}.

\bibitem{ref:stone_1992a}
        J.~M.~Stone and M.~L.~Norman, \JL{Astrophys.\ J.\ Suppl.,80,1992,753}.

\bibitem{ref:stone_1992b}
        J.~M.~Stone and M.~L.~Norman, \JL{Astrophys.\ J.\ Suppl.,80,1992,791}.

\bibitem{ref:kotake_2003}
        K.~Kotake, S.~Yamada and K.~Sato, \AJ{595,2003,304}.

\bibitem{ref:shen_1998}
        H.~Shen, H.~Toki, K.~Oyamatsu and K.~Sumiyoshi, \NPA{637,1998,435}.

\bibitem{ref:yasutake_2007}
        N.~Yasutake, K.~Kotake, M.~Hashimoto and S.~Yamada, \PRD{75,2007,084012}.


\bibitem{ref:paczynsky_wiita_1980}
		B.~Paczy\'{n}sky and P.~J.~Wiita, \JL{Astron.\ Astrophys.,88,1980,23}.


\bibitem{ref:heger_2005}
		A.~Heger, S.~E.~Woosley and H.~C.~Spruit, \AJ{626,2005,350}.

		
\bibitem{ref:balbus_Hawley_1998}
		S.~A.~Balbus and J.~F.~Hawley, \JL{Rev. Mod. Phys.,70,1998,1}.
				
\bibitem{ref:obergaulinger_2009}
		M.~Obergaulinger, P.~Cerd\'{a}-Dur\'{a}n, E.~M\"{u}ller and M.~A.~Aloy, \JL{Astron.\ Astrophys.,498,2009,241}.
		
\bibitem{ref:piran_2005}
		T.~Piran, \JL{Rev. Mod. Phys.,76,2005,1143}.
		

\bibitem{ref:granot_2004}
		J.~Granot and E.~Ramirez-Ruiz, \AJ{609,2004,L9}.
		

\bibitem{ref:nagataki_1997}
		S.~Nagataki, M.~Hashimoto, K.~Sato and S.~Yamada, \AJ{486,1997,1026}.

\bibitem{ref:nishimura_2006}
		S.~Nishimura, K.~Kotake, M.~Hashimoto, S.~Yamada, 
		N.~Nishimura, S.~Fujimoto and K.~Sato, \AJ{642,2006,410}.

\bibitem{ref:rauscher_2000}
	T.~Rauscher and F.-K.~Thielemann, \JL{At.\ Data\ Nucl.\ Data\ Tables,75,2000,1}

\bibitem{ref:rauscher_2001}	
	T.~Rauscher and F.-K.~Thielemann, \JL{At.\ Data\ Nucl.\ Data\ Tables,79,2001,47}

\bibitem{ref:goriely_2001} 
		S.~Goriely, F.~Tondeur and J.~M. Pearson, \JL{At.\ Data\ Nucl.\ Data\ Tables,77,2001,311}.
		
\bibitem{ref:meoller_1995} 
		P.~M{\"o}ller, J.~R.~Nix, W.~D.~Myers and W.~J.~Swiatecki, 
		\JL{At.\ Data\ Nucl.\ Data\ Tables,59,1995,185}.
		
\bibitem{ref:rayet_1990} 
	M.~Rayet, M.~Arnould and N.~Prantzos, \JL{Astron.\ Astrophys.,227,1990,271}.		
	

\bibitem{ref:howard_1991} 
	W.~M.~Howard, B.~S.~Meyer and S.~E.~Woosely, \AJ{373,1991,L5}.

\bibitem{ref:salpeter_1955} 
	E.~E.~Salpeter, \AJ{121,1955,161}.

\bibitem{ref:simmerer_2004} 
	J.~Simmerer, C.~Sneden, J.~J.~Cowan, J.~Collier, V.~M.~Woolf and J.~E.~Lawler, \AJ{617,2004,1091}.
	
\bibitem{ref:honda_2007} 
	S.~Honda, W.~Aoki, Y.~Ishimaru and S.~Wanajo, \AJ{666,2007,1189}.


\bibitem{ref:belic_1999} 
	D.~Belic, C.~Arlandini, J.~Besserer, J.~de Boer, J.~J.~Carroll, J.~Enders, T.~Hartmann, 
	F.~K{\"a}ppeler, H.~Kaiser, U.~Kneissl, M.~Loewe, H.~J.~Maier, H.~Maser, P.~Mohr, 
	P.~von Neumann-Cosel, A.~Nord, H.~H.~Pitz, A.~Richter, M.~Schumann, S.~Volz and A.~Zilges, 
	\PRL{83,1999,5242}.

	

		

	

	



        


		




















%










	
\end{thebibliography}
\end{document}